\documentclass[12pt]{article}
\pdfoutput=1
\usepackage[utf8x]{inputenc}
\usepackage{amssymb}
\usepackage{amsmath}
\usepackage{graphicx}
\usepackage{hyperref}

\usepackage{color}

\numberwithin{equation}{section}

\newcommand{\eq}{\begin{equation}}
\newcommand{\eqx}{\end{equation}}
\newcommand{\eqs}{\begin{equation*}}
\newcommand{\eqsx}{\end{equation*}}
\newcommand{\eqn}{\begin{eqnarray}}
\newcommand{\eqnx}{\end{eqnarray}}
\newcommand{\alg}{\begin{align}}
\newcommand{\algx}{\end{align}}

\newcommand{\f}[2]{\frac{#1}{#2}}

\newcommand{\cor}[1]{\left\langle{#1}\right\rangle}

\newcommand{\lm}{\lambda}

\renewcommand{\th}{\theta}
\newcommand{\sg}{\sigma}
\newcommand{\Sg}{\Sigma}
\newcommand{\dl}{\delta}
\newcommand{\Dl}{\Delta}
\newcommand{\al}{\alpha}
\newcommand{\bt}{\beta}
\newcommand{\om}{\omega}
\newcommand{\Om}{\Omega}
\newcommand{\gm}{\gamma}
\newcommand{\kap}{\kappa}

\newcommand{\eps}{\varepsilon}
\newcommand{\qqqq}{\quad\quad\quad\quad}

\newcommand{\tr}{\mbox{\rm tr}\,}

\newcommand{\nn}{{\cal N}}

\newcommand{\LL}{{\cal L}}

\newcommand{\Sgt}{\widetilde{\Sigma}}

\newcommand{\wb}{{\overline{w}}}
\newcommand{\ub}{{\overline{u}}}
\newcommand{\Tb}{\overline{T}}
\newcommand{\DL}{\mathbf{\Delta}}

\newcommand{\vc}[2]{\left(
\begin{matrix}
{#1} \\
{#2} \\
\end{matrix}
\right)}

\newcommand{\arr}[4]{\left(
\begin{matrix}
{#1} & {#2}\\
{#3} & {#4} \\
\end{matrix}
\right)}

\renewcommand{\ss}[2]{\left\langle #1 #2 \right\rangle}
\newcommand{\sS}[2]{\left\langle #1 \bar{#2} \right\rangle}
\newcommand{\Ss}[2]{\left\langle \bar{#1} #2 \right\rangle}
\renewcommand{\SS}[2]{\left\langle \bar{#1} \bar{#2} \right\rangle}

\newcommand{\dw}{\partial}
\newcommand{\dwb}{\bar{\partial}}
\newcommand{\gmt}{\tilde{\gm}}
\newcommand{\htl}{\tilde{h}}

\newcommand{\zeps}{\mathcal E}

\title{Correlation functions of three heavy operators 
-- the $AdS$ contribution}

\author{Romuald A. Janik\thanks{e-mail: {\tt romuald@th.if.uj.edu.pl}},\ 
Andrzej Wereszczyński\thanks{e-mail: {\tt wereszcz@th.if.uj.edu.pl}}}

\date{Institute of Physics\\
Jagiellonian University\\
ul. Reymonta 4\\ 
30-059 Krak\'ow\\
Poland}

\begin{document}

\maketitle

\begin{abstract}
We consider operators in $\nn=4$ SYM theory which are dual, 
at strong coupling,
to classical strings rotating in $S^5$. Three point correlation functions of
such operators factorize into a universal contribution coming from the $AdS$
part of the string sigma model and a state-dependent $S^5$ contribution.
Consequently a similar factorization arises for the OPE coefficients.
In this paper we evaluate the $AdS$ universal factor of the OPE coefficients
which is explicitly expressed just in terms of the anomalous dimensions
of the three operators.
\end{abstract}

\vfill
\pagebreak

\section{Introduction}

In recent years there has been huge progress in exactly solving 
the $\nn=4$ Super-Yang-Mills theory in the planar limit \cite{Bena}-\cite{ArutFrol2} . This is particularly 
remarkable as $\nn=4$ SYM is an interacting four-dimensional gauge theory 
with highly nontrivial dynamics. All previous examples of exact solvability
in Quantum Field Theories were restricted either to two dimensional theories
or, in the case of higher dimensional theories, to some supersymmetric subsector
or to theories with much simpler dynamics like topological field theories.

The solvability of $\nn=4$ SYM arises due to the AdS/CFT correspondence \cite{Mal, GubKlebPol, Wit}, 
according to which the gauge theory is equivalent to string theory in
$AdS_5 \times S^5$. Therefore $\nn=4$ SYM avoids the no-go theorems for
integrable field theories in more than two-dimensions by translating
its dynamics into properties of the two-dimensional worldsheet QFT of
the superstring in $AdS_5 \times S^5$ which is an integrable QFT.

Since $\nn=4$ SYM is a conformal field theory, all correlation functions of
local operators are, in principle, determined by a much smaller set of data:
the set of conformal dimensions (equivalently anomalous dimensions of
gauge theory operators) and the OPE coefficients. These data can be extracted
from the knowledge of 2- and 3-point correlation functions. 

In the AdS/CFT case, however, the anomalous dimensions are extracted not from
2-point correlation functions but directly as eigenvalues of the dilatation 
operator, which translates to energies of string states in $AdS_5 \times S^5$.
Therefore the problem of finding all anomalous dimensions reduces to finding
the energy levels of the two-dimensional worldsheet QFT on a cylinder.
Currently we have a very complete understanding of the spectrum of conformal 
dimensions which is described by a set of Thermodynamic Bethe Ansatz equations \cite{ArutFrol1}-\cite{ArutFrol2}.
The methods used are similar to the ones employed for relativistic integrable
field theories, although their generalization to the AdS/CFT case is far from
trivial due to many unique features of the worldsheet QFT.

For the case of OPE coefficients, however, there seems to be no alternative 
to a direct computation of 3-point correlation functions. It is convenient to 
classify the operators into three main groups, depending on their behaviour 
at strong coupling: `light', `medium' and `heavy' operators. The `light' operators
are BPS and dual to supergravity fields. Consequently their anomalous dimensions
do not depend on the coupling. The next class of operators are the lightest
massive string states whose dimensions scale like $\lm^{\f{1}{4}}$. A classical
example of these `medium' operators is the Konishi operator. Finally, the `heavy'
operators have large charges (of the order of $\lm^{\f{1}{2}}$) and are dual to
classical string states with anomalous dimensions scaling like $\lm^{\f{1}{2}}$ \cite{H_GubKlebPol}-\cite{H_Tseyt}.
Although very useful, this is only a rough and nonexhaustive classification. 
BPS operators with large
charges may for all practical purposes behave like `heavy' operators.
There may be operators with dimensions like $\lm^{\f{1}{2}}$ which may 
be very quantum and without a classical string description (like BPS operators
with two large charges).

The techniques for computing 3-point correlation functions
are very well developed for the case of `light' operators, i.e. BPS
operators dual to supergravity fields \cite{Freed}-\cite{ArutFrolBPS}. Unfortunately these OPE coefficients are
protected and do not depend on gauge theory coupling. 

For unprotected operators, the techniques for computing even 2-point correlation
functions have been developed only recently \cite{US} (but see also \cite{Tsuji}
and \cite{TseytlinB}). These results
have been used to compute OPE coefficients between two `heavy' and one `light'
operator using the known classical solution corresponding to a 2-point function
and integrating a supergravity propagator over the classical string worldsheet \cite{Zarembo}, \cite{HHL_Costa}, \cite{HHL_RoibTsey}-\cite{HHL_Ahn}. It has been further extended for correlators involving two
Wilson loops and a `light' operator \cite{AldTseyt1},\cite{AldTseyt2}

An intermediate case recently considered in the literature involved a geodesic
approximation for the three operators \cite{TristanKlose}. Such an approximation may be very
relevant to the case of three `medium' operators which are not sufficiently heavy
to generate an extended, non-pointlike, surface.

The goal of this paper is to compute (the AdS part of) the 3-point correlation 
functions of three `heavy' operators. We assume that these operators do not 
have any spin in $AdS_5$. 
The main difficulty lies in the fact that a novel type of
a classical solution has to be found. Moreover, in contrast to the spectral 
problem, there is no analog of this problem for conventional relativistic 
integrable field theories, therefore we do not have any guide in this respect.

The computation of the OPE coefficients for three `heavy' operators is 
especially interesting in view of the fundamental importance that
the integrable classification of finite-gap spinning string solutions and their
comparision with 1-loop Bethe Ansatz results had in
arriving at the all-loop interpolation. We hope that a similar comparision with
weak coupling data \cite{pert1,pert2} will be very illuminating also in the case 
of OPE coefficients.

The plan of this paper is as follows. In section~2, we briefly review the case
of 2-point correlation functions, and in section~3 the general features of the
problem of finding 3-point correlation functions. In section~4, we give an
overview of our approach to this problem, in order for the reader to
not get lost in the technicalities. In section~5, we review Pohlmeyer reduction,
and give our main technical results necessary for later computations ---
we solve functional equations for the products between the solutions
of the linear system on the 3-punctured sphere and give formulas for 
reconstructing the classical solution in $AdS_2$ from the Pohlmeyer data.
We then proceed, using these results, to evaluate in the next 3 sections the
two main parts of the AdS contribution to the correlation function. We give our
final result in section~9 and discuss the limits of small and large anomalous
dimensions and the link of the latter to the Painlev{\'e} III transcendent.
We close the paper with a discussion and several appendices with some technical
details.

\section{Two-point correlation functions}

In this section we will briefly review the computation of 2-point correlation functions
for the class of operators that we are considering in this paper, namely
operators corresponding to classical string solutions with no charges in $AdS$ \cite{H_GubKlebPol}-\cite{H_Tseyt}.

The approach introduced in \cite{US}, amounts to computing a cylinder amplitude
(with additional wavefunctions included in order to project on the specific
string state that we are interested in) 
with the boundary conditions such that the string worldsheet approaches two given
points on the boundary regularized by a cut-off $z=\zeps$. This is done by
a classical computation, where the corresponding solution is just a geodesic in
the $AdS$ part and in the $S^5$ part coincides with the unmodified $S^5$ 
spinning string solution used in
the conventional calculation of the anomalous dimension. Then one performs a saddle
point evaluation of the integral over the modular parameter. The outcome 
is\footnote{See specific examples in \cite{US}.} that the saddle point value of
the modular parameter is \emph{purely imaginary} thus effectively making the
worldsheet Euclidean.

This generic pattern indicates that we could have started directly at the saddle
point, with the worldsheet having the topology of a 2-punctured sphere (again
with small disks corresponding to $z=\zeps$ cut out), and the \emph{Euclidean} 
solution satisfying Virasoro constraints.

The Euclidean solution for the operators in question has the following simple
structure.
The $AdS_5$ part reduces just to a geodesic in the $AdS_2$ subspace which contains
the two gauge theory operator insertion points on the boundary.
Explicitly we have
\eq
\label{e.geodesic}
x = \f{R}{2}\, \tanh \kap \tau +x_0 \qqqq
z = \f{R}{2}\, \f{1}{\cosh \kap \tau}
\eqx
where the distance between the operator insertion points is $x_1-x_2=R$. Imposing
the target space cut-off $z=\zeps$ translates into a worldsheet cut-off which limits
the range of $\tau$ to
\eq
\Delta \tau= \f{2}{\kap} \log \f{R}{\zeps}
\eqx
The $S^5$ part is just the Wick rotated spinning string solution, a simple example
being a circular string with equal angular momenta given in terms of the standard
angular coordinates on $S^3 \subset S^5$ by
\eq
\psi=\sg \qqqq \phi_1=\phi_2=i \om\tau
\eqx
One should note that due to the $i$, the solution is inherently complex and in fact
does not look like any kind of spinning string. Virasoro constraints link $\kap$
and $\om$ through $\kap =\sqrt{1+\om^2}$. In general we have $\kap=\Dl$, where
$\Dl$ is the dimension of the operator.

The 2-point correlation function is now obtained by i) evaluating the $AdS_2$
action on the $AdS$ geodesic part and ii) evaluating the $S^5$ action together
with wavefunction contributions, which transforms the action integral into an
\emph{Euclidean energy} integral.

Explicitly we get
\eq
\exp\underbrace{\left\{\scriptstyle -\sqrt{\lm} \sqrt{1+\om^2} \log \f{R}{\zeps}
\right\}}_{\text{$AdS$ action}}
\cdot
\scriptstyle
\exp\underbrace{\left\{\scriptstyle -\sqrt{\lm} \sqrt{1+\om^2} \log \f{R}{\zeps}
\right\}}_{\text{$S^5$ energy}}
\eqx
which reproduces the 2-point correlation function with the correct value
of the anomalous dimension
\eq
\label{e.twopointR}
\cor{O(x_1) O(x_2)} = \left( \f{\zeps}{R} \right)^{2\sqrt{\lm} \sqrt{1+\om^2}}
\eqx
The wavefunction factors in the $S^5$ part are crucial for the correct answer. 
The Virasoro constraint
sets the only free parameter $\kap$ in the $AdS_2$ geodesic solution 
(\ref{e.geodesic}) to be equal to $\Dl$, the (anomalous) dimension of the corresponding
operator\footnote{More precisely the dimension $\DL$ is $\DL=\sqrt{\lm}\Dl$.}. Therefore,
the $AdS$ part of the solution is completely determined just by the dimension of the
operator in question and not by any further details of the specific operator. 
We will see that the same
property will hold also for 3-point correlation functions. In the following
we will use complex coordinates which, in the present case, read
\eq
w=e^{\tau+i\sg} \qqqq \wb=e^{\tau-i\sg}
\eqx 
putting the two punctures  at $w=0$ and $w=\infty$.

Finally, let us note that
the target space cut-off $\zeps$ enters (\ref{e.twopointR}) essentially as
the normalization of our operators.
This is important as in the case of 3-point correlation functions we have to retain
exactly the same normalization of operators as for 2-point functions in order to
extract unambiguosly the OPE coefficients. This leads to severe difficulties, 
such as
linking the worldsheet cut-offs around the 3 punctures to the single target space
cut-off $z=\zeps$. This is highly nontrivial due to the lack of an explicit classical
$AdS$ solution in the case of 3-point correlation functions. If one would adopt
a different approach\footnote{which should of course be in the end equivalent} of using
vertex operators \cite{TseytlinV1,Buchb,TseytlinB}, then the difficulties 
remain but appear in different places.
In the vertex operator approach, one computes the worlsheet integral over the \emph{whole}
punctured sphere with the vertex operator contributions sitting directly at the
punctures. For 2-point functions one then just neglects infinities. For 3-point
functions it is not clear how to control possible finite renormalizations. It would
be very interesting to understand quantitatively the precise dictionary 
between the two approaches.

\section{Three-point correlation functions -- general features}

In order to compute the 3-point correlation function of heavy operators, we
have to find a classical solution of string equations of motion in Euclidean 
signature with the topology of a sphere with 3 punctures, with the property that
the solution close to each puncture associated with a given gauge theory
operator $O_k$ looks asymptotically like a solution corresponding to a 2-point
correlation function of the operator $O_k$.

The classical bosonic equations of motion in $AdS_5 \times S^5$ reduce
to two independent sets of equations, one on $S^5$, the other on $AdS_5$, which are
coupled together only through the Virasoro constraint
\eq
T_{AdS_5}(w)+T_{S^5}(w)=0
\eqx
where $w$ is the holomorphic worldsheet coordinate, and $T_{AdS_5}(w)$ 
(resp. $T_{S^5}$) is the classical
energy-momentum tensor of the $AdS_5$ (resp. $S^5$) part of the $\sg$-model. 

For operators which do not have any spins, the $AdS_5$ part of the problem
greatly simplifies. Then, without loss of generality, we put the gauge theory 
operator insertion points of all three operators on a single line. Consequently, the
$AdS_5$ part of the string solution is contained in an (Euclidean) $AdS_2$ subspace. The problem,
however, does not trivialize as we are not looking for a minimal surface
but have a prescribed nonzero energy-momentum tensor $T_{AdS_2}(w)$.

Fortunately, the $AdS_2$ energy-momentum tensor can be explicitly expressed just in 
terms of the anomalous dimensions of the three operators entering the 3-point
correlation function. From now on we will denote $T_{AdS_2}(w)$ by $T(w)$.

In order to find the explicit form of $T(w)$, recall that the classical solution
should approach, at the punctures, 2-point solutions which are explicitly known.
In particular $T(w)$ for the 2-point solutions is given by
\eq
T_{2-point}(w)=\f{\Dl^2/4}{w^2}
\eqx 
Therefore at the punctures $T(w)$ should have at most poles of $2^{nd}$ order with
the leading coefficients $\Dl_k^2/4$ determined by the dimension of the operator
inserted at $w=w_k$. Since $T(w)$ is holomorphic and transforms under inversion
like a $(2,0)$ tensor
\eq
T(w) \to \f{1}{u^4} T\left( \f{1}{u} \right)
\eqx
its form is uniquely determined. Without loss of generality we may put the punctures
at $w=\pm 1$ and $w=\infty$. Then, for the case of equal conformal weights $\Dl$
at $w=\pm 1$ and $\Dl_\infty$ at $w=\infty$, $T(w)$ is given by
\eq
T(w)= \f{\Dl_\infty^2}{4} \f{w^2+a^2}{(1-w^2)^2} \qqqq 
\text{where}\quad\quad
a^2=\f{4\Dl^2}{\Dl_\infty^2}-1
\eqx
In the present paper for simplicity we will predominantly consider the above
symmetric case.

To summarize, we thus have to evaluate the action
\eq
\exp\left(-\f{\sqrt{\lm}}{\pi} \int_\Sigma \LL_{AdS_2}^{Polyakov} d^2w \right)
\equiv \left(-\f{\sqrt{\lm}}{\pi} \int_\Sigma \f{\dw z \dwb z+ \dw x \dwb x}{z^2} d^2w \right)
\eqx
for a classical solution approaching the operator insertion points $x_k$ at\\ 
$w=-1,1,\infty$
subject to the constraint
\eq
\f{(\dw z)^2 + (\dw x)^2}{z^2}=T(w)
\eqx

\section{Our strategy}

As described in the previous section, the 3-point correlation functions 
for three heavy operators with no spins in $AdS_5$ factorize into a product of
an $S^5$ and an $AdS_2$ contribution evaluated for a worldsheet with
the topology of a 3-punctured sphere. Similarly as for 2-point functions
we regularize the worldsheet by cutting out small disks of radii $\eps_i$
around the punctures which are defined by the condition that on their boundaries
\eq
z=\zeps
\eqx 
where $z$ is the $AdS$ radial coordinate in the Poincare patch ($z=0$ is the $AdS$ boundary). $\zeps$ is the target space cut-off which is taken to be very small.
It is necessary to ensure that $z=\zeps$ around each puncture in order to have
the same normalization of operators in 2- and 3-point correlation functions so as
to unambigously extract the OPE coefficients.

For the $AdS_2$ part, we have to evaluate the action of the classical solution, while
for the $S^5$ part we have to include, in addition, contributions from the classical
wavefunctions of the external states. Therefore, the 3-point correlation function
is schematically given by
\eq
\label{e.decomp}
e^{-\f{\sqrt{\lm}}{\pi} \int_{\Sigma \setminus \{\eps_i\}} \LL_{AdS_2}^{Polyakov}} \cdot 
\underbrace{\Psi_1 \Psi_2 \Psi_3^* e^{-\f{\sqrt{\lm}}{\pi} \int_{\Sigma \setminus \{\eps_i\}} 
\LL_{S^5}^{Polyakov}}}_{
e^{-\f{\sqrt{\lm}}{\pi} \int_{\Sigma \setminus \{\eps_i\}} \text{$S^5$ contribution}}
}
\eqx
Since both exponents have logarithmic divergences around the punctures, it is
convenient to subtract and add $\sqrt{T(w)\Tb(\wb)}$ regularizing the integrals.
This yields
\eq
\label{e.ads}
e^{-\f{\sqrt{\lm}}{\pi} \int_\Sigma \left( \LL_{AdS_2}^{Polyakov} -\sqrt{T\Tb} \right)
-\f{\sqrt{\lm}}{\pi} \int_{\Sigma \setminus \{\eps_i\}} \sqrt{T\Tb}}
\eqx
for the $AdS_2$ part and
\eq
\label{e.sv}
e^{-\f{\sqrt{\lm}}{\pi} \int_\Sigma \left( \text{$S^5$ contribution} -\sqrt{T\Tb} \right)
-\f{\sqrt{\lm}}{\pi} \int_{\Sigma \setminus \{\eps_i\}} \sqrt{T\Tb}}
\eqx
for the $S^5$ part. The first terms in the above expressions are now finite and can
be integrated over the whole punctured sphere, while the explicit dependence on the
worldsheet cut-offs appears only in the second integral with a known integrand.

In this paper we will compute the contribution (\ref{e.ads}) together with the
\emph{second} term in (\ref{e.sv}), leaving the remaining factor
\eq
\label{e.svii}
e^{-\f{\sqrt{\lm}}{\pi} \int_\Sigma \left( \text{$S^5$ contribution} -\sqrt{T\Tb} \right)}
\eqx
for further investigation.

In order to compute 
\eq
\label{e.polregi}
e^{-\f{\sqrt{\lm}}{\pi} \int_\Sigma \left( \LL_{AdS_2}^{Polyakov} -\sqrt{T\Tb} \right)}
\eqx
we will use Pohlmeyer reduction \cite{Pohl}, \cite{deVega} and adapt the methods of \cite{AldMal} to evaluate this expression.
Firstly, one transforms the above integral into an integral of the wedge product
of two closed 1-forms on a double cover of $\Sigma$. Secondly, one uses Riemann reciprocity
(Riemann bilinear identity) to express the integral in terms of products of integrals
of the 1-forms on certain open cycles. Thirdly, one links the above 1-form integrals
to the asymptotics in the spectral parameter ($\xi \to 0$) of appropriate skew products between
specific solutions (associated with each puncture) of the Pohlmeyer linear system.
Thus the evaluation of the integral (\ref{e.polregi}) is reduced to the knowledge
of appropriate skew products as a function of the spectral parameter.

The remaining integral 
\eq
\label{e.divint}
e^{-\f{\sqrt{\lm}}{\pi} \int_{\Sigma \setminus \{\eps_i\}} \sqrt{T\Tb}}
\eqx
can be evaluated analytically in the small $\eps_i$ limit. The main difficulty lies
in linking the worldsheet cut-offs $\{\eps_i\}$ to the target space cut-off $z=\zeps$,
without an explicit knowledge of the classical solution. To do that we need formulas
for reconstructing the classical solution from Pohlmeyer data. Fortunately, since
the classical solution should approach the known solutions for 2-point functions
close to the punctures, we can get explicit formulas relating 
the positions of the
gauge theory operator insertion points $x_k$ and the target space cut-off $\zeps$
to the worldsheet cut-offs $\{\eps_i\}$ in terms of the skew products mentioned
above, but this time evaluated at $\xi=1$. Using this knowledge, the (two copies of the) 
integral (\ref{e.divint}) yield the standard space-time dependent part of the
3-point CFT correlation function, as well as a finite contribution to the OPE
coefficient expressed in terms of the skew products at $\xi=1$.

The skew products between the specific solutions (of the Pohlmeyer linear system)
associated to each puncture 
are therefore a key ingredient in the evaluation
of the 3-point correlation function. We will often refer to these 
chosen solutions as `elementary solutions'. 

In the following section we introduce the main
features of Pohlmeyer reduction, the elementary solutions associated with each puncture
and define the skew products. Then we derive and solve functional equations for
the skew products of the elementary solutions as a function of the spectral parameter $\xi$.
Finally we state the reconstruction formulas which link the operator insertion points
and the target space cut-off to appropriate skew products.

After this preparatory part, we proceed to evaluate the integral (\ref{e.polregi})
in section~7 and the divergent contribution (\ref{e.divint}) in section~8.
Then we put together the obtained formulas into the final AdS contribution 
to the OPE coefficient and analyze the limits of large and small anomalous 
dimensions as well as the extremal limit.

Before we end this overview, let us remark that the same decomposition (\ref{e.decomp})
could also be performed for a 2-punctured sphere corresponding to a 2-point
correlation function (of course in this case only two wavefunctions would appear). Then it turns out that both the `nontrivial' parts (\ref{e.polregi}) and (\ref{e.svii})
are identically zero. However it is interesting to note that they vanish for quite
different reasons. The AdS part (\ref{e.polregi}) vanishes because it is evaluated
on a trivial classical solution -- a point-like string moving along a geodesic.
The corresponding Pohlmeyer function is just identically zero and consequently
(\ref{e.polregi}) vanishes. On the $S^5$, however, we deal with arbitrarily complicated
finite-gap solutions of arbitrary genus, which would have a highly nontrivial Pohlmeyer
description. Yet, the wavefunction contributions transform the classical action
into an integral of the energy density (in an appropriate coordinate system)
and the resulting $S^5$ contribution exactly cancels the integral of $\sqrt{T\Tb}$. It is
tempting to speculate that a similar simplification may occur for the case of
3-point functions.

\section{Pohlmeyer reduction}

Contrary to the well known case of Pohlmeyer reduction for minimal surfaces in 
$AdS_3$ \cite{Pohl}, \cite{deVega}, \cite{AldMal}, we need to perform Pohlmeyer reduction for classical solutions in $AdS_2$ but
with a prescribed nonzero energy-momentum tensor. Thus the classical solutions in $AdS_2$ are of course
\emph{not} minimal surfaces. On the other hand, the full string solution, which takes into account both $AdS_2$ and $S^5$ contributions, is a minimal surface.

The Pohlmeyer reduction for this case amounts to defining 
$\gmt(w,\wb)$ through
\eq
\label{e.pohl1}
\f{\dw x \dwb x+ \dw z \dwb z}{z^2}= \sqrt{T\Tb} \cosh \gmt
\eqx
where $T$ is the energy-momentum tensor $T(w)$.
Then $\gmt(w,\wb)$ satisfies a modified form of Sinh-Gordon equation
\eq
\label{e.pohl1eom}
\dw \dwb \gmt= \sqrt{T\Tb} \sinh \gmt
\eqx
The solution corresponding to a 2-point function is just $\gmt(w,\wb)\equiv 0$.
Consequently, the boundary conditions close to each puncture are
\eq
\gmt \to 0
\eqx
For the case relevant to 3-point correlation functions, $T(w)$ has two zeroes, and
thus the form of Pohlmeyer reduction given by (\ref{e.pohl1}) is inconvenient as
it would imply that all first derivatives vanish
\eq
\dw z=\dwb z=\dw x=\dwb x=0
\eqx 
at the zeroes of $T(w)$. This would be a very nongeneric situation as each 
such single equation gives a codimension one subspace. Their intersection
is generically empty. This is even the case for pointlike strings appearing
in 2-point functions. Consequently we will assume, 
as is the case for polygonal Wilson loops, that the right hand side of (\ref{e.pohl1}) 
is everywhere nonzero. This implies that $\gmt$ has to have logarithmic 
singularities at the zeros of $T(w)$. 

To avoid this drawback, it is convenient to redefine $\gmt$ through
\eq
\label{e.gmt}
\gmt =2\gm-\f{1}{2} \log T\Tb
\eqx
Now (\ref{e.pohl1}) takes the form
\eq
\f{\dw x \dwb x+ \dw z \dwb z}{z^2}= \f{1}{2} \left( e^{2\gm}+ T\Tb e^{-2\gm} \right)
\eqx
which does not lead to any problem at the zeroes of $T(w)$.
The equation of motion becomes
\eq
\label{e.mshgeom}
\dw \dwb \gm=\f{1}{4} \left( e^{2\gm}- T\Tb e^{-2\gm} \right)
\eqx
This is virtually the same as the setup for Wilson loop \cite{AldMal}, \cite{AMSV} but with the polynomial
defining the polygonal Wilson loop substituted by $T(w)$. 
We will discuss the similarities and differences in more detail at the end
of the present section.

It is well known that the modified sinh-Gordon model is integrable.
It is easiest to verify by making a holomorphic change of worldsheet 
coordinates to map this model into ordinary sinh-Gordon.
However, due to the rather complicated analytical structure 
of the resulting domain we will not use this mapping in the sequel.

Below we review the main features of the integrability of sinh-Gordon model which
will be important for us later. 
 
There exists a family of flat connections parametrized by the spectral parameter
$\xi$. We will also use the parametrization
\eq
\xi = e^\th
\eqx
The flat connection $J=J_w\, dw+J_\wb\, d\wb$ has the following components
\eq
\label{e.flatcon}
J_w= \f{1}{2} \arr{\dw \gm}{-\f{1}{\xi} e^\gm \;}{-\f{1}{\xi} e^{-\gm} T}{-\dw \gm}
\qqqq
J_{\wb}=\f{1}{2} \arr{-\dwb \gm}{-\xi e^{-\gm} \Tb \;}{-\xi e^\gm}{\dwb \gm}
\eqx
Flatness is equivalent to the compatibility of the associated linear system
\eq
\label{e.linear}
\dw \Psi+J_w \Psi =0 \qqqq \dwb \Psi+J_{\wb} \Psi =0
\eqx
which in turn is equivalent to the equation of motion (\ref{e.mshgeom}).
Another useful decomposition of the flat connection is 
\eq
J=\f{1}{\xi}\, \Phi_w\, dw +A+ \xi\, \Phi_{\wb}\, d\wb
\eqx
using which we may write the string action as
\eq
\f{\dw x \dwb x+ \dw z \dwb z}{z^2}= 2\, \tr \Phi_w \Phi_{\wb}
\eqx

Certain specific solutions of the linear system (\ref{e.linear}) associated with 
each puncture will be of key importance in the following. Since close to the punctures
\eq
T(w) \sim \f{\Dl^2/4}{w^2} 
\eqx 
and
\eq
\gm \sim \f{1}{4} \log T(w) \Tb(\wb)
\eqx
we get two solutions, which close to the puncture behave like
\eq
\tilde{\Psi}_1= w^{\f{\Dl}{4 \xi}} \wb^{\f{\Dl}{4} \xi} 
  \vc{w^{\f{1}{4}} \wb^{-\f{1}{4}}}{w^{-\f{1}{4}} \wb^{\f{1}{4}}}
\qqqq
\tilde{\Psi}_2= w^{-\f{\Dl}{4 \xi}} \wb^{-\f{\Dl}{4} \xi} 
  \vc{w^{\f{1}{4}} \wb^{-\f{1}{4}}}{-w^{-\f{1}{4}} \wb^{\f{1}{4}}}
\eqx
It is clear that these solutions have nontrivial monodromies $e^{\pm i \tilde{p}(\xi)}$
around the puncture $w=0$ with 
\eq
\tilde{p}(\xi)=\Dl \f{\pi}{2} \left( \xi -\f{1}{\xi} \right)+\pi
\eqx
It is in fact convenient to get rid of the $\pi$ by a gauge transformation 
$\Psi=V \tilde{\Psi}$ with
\eq
\label{e.vgauge}
V=\arr{\left(\f{(w-w_1)(w-w_2)(w-w_3)}{(\wb-\wb_1)(\wb-\wb_2)(\wb-\wb_3)}\right)^{-\f{1}{4}} }{0}{0}{
\left(\f{(w-w_1)(w-w_2)(w-w_3)}{(\wb-\wb_1)(\wb-\wb_2)(\wb-\wb_3)}\right)^{\f{1}{4}} }
\eqx
Then our final basis of solutions associated to the puncture at $w=w_1$ is
\eqn
\label{e.assol1}
\Psi_1 &=& \f{i}{\sqrt{2}} (w-w_1)^{\f{\Dl}{4 \xi}} (\wb-\wb_1)^{\f{\Dl}{4} \xi}
\vc{u_1}{u_1^{-1}}  \\
\label{e.assol2}
\Psi_{\bar{1}} &=& \f{i}{\sqrt{2}} (w-w_1)^{-\f{\Dl}{4 \xi}} 
(\wb-\wb_1)^{-\f{\Dl}{4} \xi}
\vc{u_1}{-u_1^{-1}} 
\eqnx
where the constants $u_1$ are given by
\eq
u_1=\f{(\wb_{12}\wb_{13})^{\f{1}{4}}}{(w_{12} w_{13})^{\f{1}{4}}}
\eqx
with $w_{ij}=w_i-w_j$. The solutions $1$ (i.e. $\Psi_1$) and $\bar{1}$ (i.e. $\Psi_{\bar{1}}$) 
have the monodromies $e^{ip(\xi)}$ and $e^{-ip(\xi)}$
with the pseudomomentum given by
\eq
\label{e.pseudo}
p(\xi)=\Dl \f{\pi}{2} \left( \xi -\f{1}{\xi} \right) \quad (\equiv \Dl \pi \sinh \th)
\eqx
Several comments are in order here. These solutions can be continued to the
neighborhoods of the other punctures. Of course we do not know their analytical expressions
so we cannot perform this explicitly. Generically these solutions will no longer be
eigenstates of the monodromy operator around \emph{other} punctures. However, since
the space of solution of the linear system is two-dimensional, we can express $1$ and
$\bar{1}$ as linear combinations\footnote{with coefficients depending just on the spectral
parameter} of an analogous basis $k$ and $\bar{k}$ at the puncture $w=w_k$. 
It is exactly these coefficients which are the key ingredients for the computation of the
AdS part of the 3-point correlation function.
In order
to fix an inherent ambiguity associated with nontrivial monodromy, we have to fix
once and for all the path of analytical continuation, whose detailed form will not
be important for us.

It is clear that the pseudomomentum of the elementary solutions obeys the important 
general property
\eq
p(e^{i\pi} \xi)=-p(\xi)
\eqx  
This suggests that it should be possible to obtain the second solution $\bar{1}$ from
the first $1$. Since just changing $\xi \to e^{i\pi} \xi$ modifies the expressions
for the flat connection, one has to perform in addition a similarity transformation
\eq
U J_{w,\wb}(w,\wb;\xi) U^{-1}=J_{w,\wb}(w,\wb;e^{i\pi}\xi)
\eqx
with $U=i\sg_3$ to compensate. Therefore the second solution can be obtained from
the first $\Psi(w,\wb;\xi)$ through
\eq
\label{e.sg3prop}
\Psi_{\bar{k}}(w,\wb;\xi)=\sg_3 \Psi_k(w,\wb;e^{i\pi}\xi)
\eqx
This is a crucial property which allows for the formulation of a set of functional 
equations for the overlap coefficients.  

Let us close this section with a comparision of the present set-up of a 3-point
correlation function in $AdS_2$ with the case of Pohlmeyer reduction 
for polygonal Wilson loops in $AdS_3$.

In both cases we have the same modified sinh-Gordon model, but with the modification
defined in terms of functions with quite different analytical properties. In the case of
polygonal Wilson loops we have a polynomial with a single asymptotic region (covered by
several Stokes sectors), here we have a rational function with three ($2^{nd}$ order) poles
and thus we have three distinct asymptotic regions. In the Wilson loop case, only
the `small' solution was unambigously defined, while here two solutions are uniquely specified
as eigenfunctions of the monodromy operator at each puncture. Finally, the spacetime
picture is quite different. In the Wilson loop case, the target-space was $AdS_3$ and
one had natural `left-' and `right-' linear problems. Here the target space is
one dimension less ($AdS_2$) and we have to develop appropriate reconstruction
formulas and impose boundary conditions characteristic of a 3-point correlation
function (i.e. fixing the boundary coordinates of the operator insertion 
points $x_k$ and the target-space cut-off $z=\zeps$).

\subsection{Overlaps}

In this section we will derive and solve functional equations for the overlaps
between the elementary solutions associated with each puncture defined in the previous 
section. For any two solutions of the linear system (\ref{e.linear}) $\Psi_k$ and $\Psi_l$, 
one defines the antisymmetric product (skew-product)
\eq
\ss{k}{l}
\eqx
which is the determinant of the matrix formed by the column vectors $\Psi_k$ 
and $\Psi_l$. It is a function of the spectral parameter $\xi$ (or equivalently $\th$). 
Our elementary solutions (\ref{e.assol1})-(\ref{e.assol2}) have 
the canonical normalization
\eq
\sS{k}{k}=1
\eqx
A characteristic feature of the product $\ss{k}{l}$ is that for \emph{any} four
solutions the relevant products obey a purely algebraic relation called the Schouten
identity:
\eq
\ss{i}{j}\ss{k}{l}+\ss{i}{l}\ss{j}{k}+\ss{i}{k}\ss{l}{j}=0
\eqx

In our case we have 6 distinguished solutions of the linear system -- 
$1,\bar{1},2,\bar{2},3,\bar{3}$. Our aim is to find the skew-products between
these solutions as functions of $\th$, given the set of conformal weights $\Dl_1$,
$\Dl_2$ and $\Dl_3$.

It is convienient to repackage the products between the various solutions into 
connection matrices $M_{kl}$ which transform the coordinates of a solution 
in the basis associated to the puncture $l$ to the coordinates in the basis 
associated to the puncture $k$.

The equation
\eq
\vc{\gm}{\dl} =\underbrace{\arr{A}{B}{C}{D}}_{M_{kl}} \vc{\al}{\bt}
\eqx
amounts to the following equality between the elementary solutions
\eq
\gm \Psi_k+\dl \Psi_{\bar{k}}= \al \Psi_l+\bt \Psi_{\bar{l}}
\eqx
This means that
\eqn
\Psi_l &=& A \Psi_k+C \Psi_{\bar{k}} \\
\Psi_{\bar{l}} &=& B \Psi_k+D \Psi_{\bar{k}}
\eqnx
Now taking appropriate products gives an expression for $M_{kl}$ in terms of our fundamental products.
\eq
M_{kl}=\arr{-\Ss{k}{l}}{-\SS{k}{l}}{\ss{k}{l}}{\sS{k}{l}}
\eqx
The obvious compatibility conditions between the connection matrices
\eq
\label{e.compat}
M_{km}=M_{kl} M_{lm} \qqqq M_{kl} M_{lk}=id
\eqx
are in fact \emph{equivalent} to the full set of Schouten identities. This can be easily seen by considering various choices for the solutions entering Schouten identities and comparing with appropriate elements
of the matrix products (\ref{e.compat}).

The full set of functional relations for the products $\ss{k}{l}$ thus comprises the compatibility conditions (\ref{e.compat}) and the vanishing of total monodromy
\eq
\label{e.monodromyeq}
\Om_1 M_{13} \Om_3 M_{32} \Om_2 M_{21}=id
\eqx
where
\eq
\Om_k=\arr{e^{ip_k(\th)}}{0}{0}{e^{-ip_k(\th)}}
\eqx

As they stand, the equations (\ref{e.compat}) and (\ref{e.monodromyeq}) are 
a complicated set of constraints for 12 unknown products. The key property which
allows us to transform them into a set of solvable functional equations is the 
property (\ref{e.sg3prop}). Using this construction we may relate 6 of the 12 unknown
products to the other 6 but evaluated at a shifted value of the spectral parameter.
Explicitly suppose that the $k$ elementary solution (eigenvector of the monodromy
matrix around the puncture $w_k$ with the eigenvalue $e^{ip_k}$) is
\eq
\Psi_k(w,\wb;\xi)=\vc{a_k}{b_k}
\eqx
Then the second solution $\bar{k}$ (with eigenvalue $e^{-ip_k}$) is obtained through
\eq
\Psi_{\bar{k}}(w,\wb;\xi)=\sg_3 \Psi_k(w,\wb;\xi e^{i\pi})\equiv 
\vc{a_k^{++}}{-b_k^{++}}
\eqx
where the superscript `$+$' denotes the shift $\th \to \th+i\pi/2$. Using 
the fact\footnote{This follows from $\Psi_k^{++++}=\lm_k \Psi_k$ and an argument that
$\lm_k=1$.} that $a_k^{++++}=a_k$ and $b_k^{++++}=b_k$, it follows that
\eqn
\SS{k}{l}&=& -\ss{k}{l}^{++} \\
\Ss{k}{l}&=& -\sS{k}{l}^{++}
\eqnx
Now the relations (\ref{e.compat}) and (\ref{e.monodromyeq}) become functional
equations for just 6 products.

\subsubsection*{Solution of the functional relation}

In this section we will solve the full set of functional equations (\ref{e.compat}) 
and (\ref{e.monodromyeq}). 
Let us first define the three functions
\eqn
X_{32} &\equiv& \ss{3}{2} \ss{3}{2}^{++} \\
X_{3\bar{2}} &\equiv& \sS{3}{2} \sS{3}{2}^{++} \\
X_{2\bar{1}} &\equiv& \sS{2}{1} \sS{2}{1}^{++}
\eqnx
Once we determine them explicitly, the products $\ss{3}{2}$, 
$\sS{3}{2}$ and $\sS{2}{1}$ will be expressed through convolution
with a $\cosh$ kernel and zero-mode parts. The remaining products $\ss{2}{1}$, 
$\ss{3}{1}$ and $\sS{3}{1}$ will turn out to be expressed in terms of 
the first three and the pseudomomenta. In fact for our applications, it suffices
to know the formulas for the products between the unbarred solutions: $\ss{1}{2}$,
$\ss{2}{3}$ and $\ss{3}{1}$ -- so all of them will be expressed through $X_{32}$ and
some permutation of indices. However in this section, for completeness, 
we will solve all equations.

We start from the equation $M_{32}M_{21}=M_{31}$. This just expresses 
$\ss{3}{1}$ and $\sS{3}{1}$ through the Schouten identities:
\eqn
\ss{3}{1} &=& \ss{3}{2} \sS{2}{1}^{++}+\ss{2}{1}\sS{3}{2} \\
\sS{3}{1} &=& \ss{3}{2} \ss{2}{1}^{++}+\sS{2}{1}\sS{3}{2}
\eqnx

Then define $Y_1$ and $Y_3$ as
\eqn
Y_1 &=& \f{\sS{3}{2} \ss{2}{1}}{\ss{3}{1}} \\
Y_3 &=& \f{\sS{3}{2} \sS{2}{1}}{\sS{3}{1}}
\eqnx
We can express $\ss{2}{1}$ in terms of $Y_1$. Plugging the results into 
the formula for $Y_3$ we see that we can express $X_{3\bar{2}}$ as
\eq
X_{3\bar{2}} = X_{32} \f{Y_1^{++} Y_3}{(1-Y_1^{++})(1-Y_3)} 
\eqx

At this stage is convenient to rewrite the monodromy equation (\ref{e.monodromyeq})
in the form
\eq
\label{e.monodromy}
M_{32} \Om_2 M_{21} = \Om_3^{-1} M_{31} \Om_1^{-1}
\eqx
The entries of (\ref{e.monodromy}) are in fact the counterparts of the $\bar{Y}$ 
functions introduced by Maldacena and Zhiboedov (\cite{MaldZhib} fig. 5).

The equation (\ref{e.monodromy}) enables us to determine $Y_1$ and $Y_3$ defined
earlier. Explicitly we obtain
\eqn
Y_1 &=& \f{1-e^{i(p_3-p_1-p_2)}}{e^{2ip_2}-e^{-2ip_2}} \\
Y_3 &=& \f{1-e^{i(p_3+p_1-p_2)}}{e^{2ip_2}-e^{-2ip_2}}
\eqnx

Now we proceed to the equation $M_{21} M_{13}=M_{23}$ which enables us to 
determine $X_{2\bar{1}}$, and finally the equation $M_{13} M_{32}=M_{12}$ 
determines $X_{32}$. We check that all the remaining compatibility conditions
of (\ref{e.compat}) are satisfied.

Consequently, the final functional equations for $\ss{3}{2}$, $\sS{3}{2}$ 
and $\sS{2}{1}$ read
\eqn
\ss{3}{2} \ss{3}{2}^{++} &=& \f{\sin \f{p_1-p_2-p_3}{2} \sin \f{p_1+p_2+p_3}{2}}{\sin p_2 \sin p_3} \\
\sS{3}{2} \sS{3}{2}^{++} &=& \f{\sin \f{p_3-p_1-p_2}{2} \sin \f{p_2-p_1-p_3}{2}}{\sin p_2 \sin p_3} \\
\sS{2}{1} \sS{2}{1}^{++} &=& \f{\sin \f{p_1-p_2-p_3}{2} \sin \f{p_2-p_1-p_3}{2}}{\sin p_1 \sin p_2} 
\eqnx
with the right hand sides being exactly our functions $X_{32}$, $X_{3\bar{2}}$ and
$X_{2\bar{1}}$. In the above expressions we did not use the specific form of $p_k(\th)$
given by (\ref{e.pseudo}) but only the generic property
\eq
p_k(\th+ i \pi)=-p_k(\th)
\eqx
Therefore, the above solution may have a much greater range of applicability
than the specific case of no spin in $AdS_5$ that we consider in the present paper.

Let us now specialize to the pseudomomenta (\ref{e.pseudo}) and use the parametrization $\xi=e^\th$. Then
\eq
p_k(\th)=\Dl_k \pi \sinh \th
\eqx
The above functional equations can be recast in the form
\eqn
\ss{3}{2}^+ \ss{3}{2}^{-} &=& -\f{\sinh(\f{\Dl_2+\Dl_3-\Dl_1}{2} \pi \cosh\th) \sinh(\f{\Dl_1+\Dl_2+\Dl_3}{2} \pi \cosh\th)}{
\sinh( \Dl_2 \pi \cosh\th ) \sinh( \Dl_3 \pi \cosh\th ) } \\
\label{eq2}
\sS{3}{2}^+ \sS{3}{2}^{-} &=& \f{\sinh(\f{\Dl_1+\Dl_3-\Dl_2}{2} \pi \cosh\th) \sinh(\f{\Dl_1+\Dl_2-\Dl_3}{2} \pi \cosh\th)}{
\sinh( \Dl_2 \pi \cosh\th ) \sinh( \Dl_3 \pi \cosh\th ) } \\
\label{eq3}
\sS{2}{1}^+ \sS{2}{1}^{-} &=& \f{\sinh(\f{\Dl_2+\Dl_3-\Dl_1}{2} \pi \cosh\th) \sinh(\f{\Dl_1+\Dl_3-\Dl_2}{2} \pi \cosh\th)}{
\sinh( \Dl_1 \pi \cosh\th ) \sinh( \Dl_2 \pi \cosh\th ) }
\eqnx

As mentioned before, for our purposes we will be interested in the solution of the
first equation. The formulas for $\ss{1}{2}$ and $\ss{3}{1}$ can then be obtained 
simply by a permutation of the $\Dl_i$'s.

The right hand side of the first equation has the property that it approaches
a constant when $\th \to \pm \infty$ thus making the solution simpler.
The basic functional equation to solve is
\eq
\label{e.basic}
f_a^+ f_a^- =1-e^{-a \pi \cosh\th}
\eqx
which can be solved by convolution
\eq
f_a(\th)=\exp \int_{-\infty}^\infty \f{d\th'}{2\pi} \f{\log\left( 1-e^{-a \pi 
\cosh\th'} \right)}{ \cosh( \th -\th')}
\eqx
Therefore we get the following expression for the product $\ss{3}{2}$:
\eq
\label{e.s32}
\ss{3}{2}(\th)=i e^{M e^{\th} +M^* e^{-\th}} \cdot 
\f{f_{\Dl_2+\Dl_3-\Dl_1}(\th) f_{\Dl_2+\Dl_3+\Dl_1}(\th)}{f_{2\Dl_2}(\th)f_{2\Dl_3}(\th)}
\eqx
where the first term is a zero-mode part depending on two constants $M$ and $M^*$.
These constants can be found from the leading WKB asymptotics of $\ss{3}{2}(\th)$
which can be found independently. We will discuss this part in section~8 and
Appendix~C.

The formula (\ref{e.s32}) is the key formula of this section. We will use it in
the following to obtain the AdS contribution to the 3-point correlation functions. 

Before we end this section, for completeness, let us discuss briefly 
the solution of equations (\ref{eq2})
and (\ref{eq3}). The right hand sides of these equations do not approach a constant
when $\th \to \pm \infty$ so we cannot directly use the convolution with the $\cosh$
kernel. Apart from (\ref{e.basic}) we just have to consider in addition
\eq
\tilde{f}_a^+ \tilde{f}_a^- =e^{\f{a}{2} \pi \cosh \th}
\eqx
with the solution
\eq
\tilde{f}_a(\th)=e^{-\f{a}{2} \th \sinh\th}
\eqx
This will then solve the functional equations for $\sS{3}{2}$ and $\sS{2}{1}$.
However in this case the zero mode part is undetermined. We will not consider this
issue further since we do not need these expressions in the remaining part of the paper.

\subsection{Reconstruction formulas}

In this section we will show how one can reconstruct the string solution in
the $AdS_2$ target space from the Pohlmeyer data.
The explicit expressions for the string solutions are important for two reasons. 
Firstly, the correlation function has to be regularized by making a cut-off at $z=\zeps$. 
This has to be translated into a worldsheet cut-off around 
each puncture. Secondly, we need to have control over the coordinates of 
the operator insertion points $x_k$ in gauge theory. In particular the standard 
conformal dependence on $x_k$ should arise automatically.

We will show that the string solution can be reconstructed from \emph{two} given 
solutions $\Psi_A$ and $\Psi_B$ of the linear system for $\th=0$ ($\xi=1$) normalized by 
$\ss{\Psi_A}{\Psi_B}=1$. Equivalently, it is determined by the coefficients
$\al$, $\bt$, $\gm$ and $\dl$ of
\eq
\label{e.phi12}
\Psi_A=\al \Psi_1 +\bt \Psi_{\bar{1}} \qqqq \Psi_B=\gm \Psi_1 +\dl \Psi_{\bar{1}}
\eqx
satisfying $\al \dl-\bt \gm=1$. These two solutions can also be combined 
into a $2\times 2$ matrix as
\eq
\hat{\Psi}= \left( \Psi_A \Psi_B \right) \equiv \arr{a}{b}{c}{d}
\eqx

\subsubsection*{Global embedding coordinates}

We will present reconstruction formulas in the global embedding coordinates
\eq
Y^1 = \f{-1}{2z} (1-x^2-z^2) \quad\quad
Y^2 = \f{1}{2z} (1+x^2+z^2) \quad\quad
Y^3 = \f{x}{z}
\eqx
satisfying
\eq
(Y^1)^2-(Y^2)^2+(Y^3)^2=-1 \quad\quad 
(dY^1)^2-(dY^2)^2+(dY^3)^2=\f{dx^2+dz^2}{z^2}
\eqx
Once we know $Y^i$, the Poincare coordinates may be easily extracted through
\eq
Y^2-Y^1=\f{1}{z} \qqqq Y^3=\f{x}{z}
\eqx 

\subsubsection*{Reconstruction formulas}

The reconstruction formula for the string solution is
\eq
\label{e.YIrecon}
Y^I=\f{1}{2} \tr\left( \tilde{\sg}^I C \hat{\Psi}^T D \hat{\Psi} \right)
\eqx 
where
\eq
C =\arr{0}{1}{-1}{0} \qqqq D=\arr{0}{i}{i}{0}
\eqx
and $\tilde{\sg}$ are related to the standard Pauli matrices by
\eq
\tilde{\sg}^1=\sg^1 \quad \quad
\tilde{\sg}^2=i\sg^2 \quad \quad
\tilde{\sg}^3=\sg^3
\eqx
Using the equations
\eq
\dw \hat{\Psi}+J \hat{\Psi}=0 \qqqq \dwb \hat{\Psi}+\bar{J} \hat{\Psi}=0
\eqx
written in the original gauge (\ref{e.flatcon}), we may verify that
\eqn
(Y^I)^2 &=& -1 \\
(\dw Y^I)^2 &=& T(w) \\
(\dwb Y^I)^2 &=& \Tb(\wb) \\
(\dw Y^I \dwb Y^I) &=& \f{1}{2} \left( e^{2\gm}+T\Tb e^{-2\gm} \right) \\
\dw \dwb Y^I &=& (\dw Y^K \dwb Y^K) Y^I
\eqnx

From the formula (\ref{e.YIrecon}) we may now express the $AdS_2$ coordinates
directly in terms of the components of $\hat{\Psi}$ given above:
\eq
\label{e.recon}
\f{1}{z} \equiv Y^2-Y^1 =2iac \qqqq
\f{x}{z} \equiv Y^3 =i(ad+bc)
\eqx
Note that these expressions are invariant under the gauge transformation
\eq
\Psi \to \arr{\lm}{0}{0}{\lm^{-1}} \Psi
\eqx
thus we can use them also in our final gauge (\ref{e.vgauge}).

\subsubsection*{The operator insertion points $x_k$ and the target space cut-off $\zeps$}

We may now use the above formulae to express the gauge theory operator insertion
points $x_k$ in terms of the two solutions of the linear system $\Psi_A$, $\Psi_B$ which determine
the classical string embedding. Fortunately, close to the puncture we have 
explicit formulas (\ref{e.assol1})-(\ref{e.assol2}) for the basis of solutions
around each puncture. Using these formulas we see that for $\xi=1$, the dominant 
solution around the puncture $w_k$ is $\bar{k}$.
So only the $\bt$ and $\dl$ coefficients of $\Psi_{A,B}$ in 
(\ref{e.phi12}) will be relevant.

Using the formulas (\ref{e.recon}) and the explicit expression 
(\ref{e.assol2}) we get the link between target space $z$ coordinate and 
the worldsheet coordinate around the puncture $w=w_k$
\eq
z=\f{1}{i\bt_k^2} |w-w_k|^{\Dl_k} \quad\quad \text{where} \quad
\bt_k =\ss{k}{\Psi_A}
\eqx
This allows us to relate the target space cut-off $z=\zeps$ to
the worldsheet cut-offs $\eps_k$:
\eq
\label{e.epsk}
\Dl_k \log \eps_k =\log \zeps + \log |\ss{k}{\Psi_A}|^2
\eqx
Similarly, we obtain expressions for the coordinates of the gauge theory
operator insertion points
\eq
\label{e.xk}
x_k=\f{\ss{k}{\Psi_B}}{\ss{k}{\Psi_A}}
\eqx

The two above expressions (\ref{e.epsk}) and (\ref{e.xk}) are the key results
of the present section which will be essential for the determination of
the `divergent' part of the AdS action integral in section~8 (recall also 
the overview in section~4 above).

\section{The AdS action}

After the above preparations we are now ready to tackle the calculation
of the AdS contribution to the 3-point correlation function using Pohlmeyer
reduction. 

We have to compute the action of the $AdS_2$ part of the solution over the 
worldsheet, which is a `regularized' 3-punctured sphere with 3 disks cut out
around the punctures\footnote{For the puncture at $w=\infty$ we define the worldsheet
cut-off through $|w|<1/\eps_\infty$.} $|w-w_i|>\eps_i$. The worldsheet cut-off's
around each puncture are not independent but are determined by 
the \emph{single} target-space cut-off $z=\zeps$
\eq
\f{\sqrt{\lm}}{\pi}\int_{\Sigma \setminus \{\eps_i\}} \f{\dw x \dwb x+ \dw z \dwb z}{z^2}
\eqx
Let us emphasize that this is \emph{not} the area of the worldsheet as there
is a nonzero energy-momentum tensor. Using the elements of the Pohlmeyer flat 
connection the above integral can be written as
\eq
\f{\sqrt{\lm}}{\pi}\int_{\Sigma \setminus \{\eps_i\}} 2\, \tr \Phi_w \Phi_\wb
\eqx 
Since in the above expression we have both an unknown integrand (i.e. which depends
on the solution of the modified sinh-Gordon equation which we do not know explicitly)
and an unknown integration domain (since the worldsheet cut-offs depend on the
target-space solution), it is convenient, as outlined in section~4, to split 
the integral into a cut-off independent finite piece which involves the unknown
integrand but can be integrated over the whole punctured sphere and a cut-off
dependent part with an explicitly known integrand.
\eq
\label{e.decomp2}
\f{\sqrt{\lm}}{\pi}\int_{\Sigma} \left( 2\, \tr {\Phi_w \Phi_\wb}-
\sqrt{T \Tb} \,d^2w\right) + 
\f{\sqrt{\lm}}{\pi}\int_{\Sigma \setminus \{\eps_i\}} \sqrt{T \Tb}\,d^2w
\eqx 
As outlined in section~4, we also adopt a similar regularization for the $S^5$
part thus the cut-off dependent part will appear in the final answer with
coefficient 2.
We will evaluate the first integral in section 7, and the second integral 
in section 8.

\section{The regularized Pohlmeyer contribution}

In order to evaluate the first integral in (\ref{e.decomp2}), we will proceed as for
Polygonal Wilson loops, and pass to a gauge\footnote{By this we mean redefining
the solution of the linear system $\tilde{\Psi}=W\Psi$ with some given matrix 
$W$ depending on the worldsheet coordinates $w$, $\wb$.} where the $\Phi_w$ part of the flat
connection is diagonal
\eq
\Phi_w \to W \Phi_w W^{-1}
\eqx
Fortunately it turns out that the diagonalized $\Phi_w$ does not depend on the
unknown Pohlmeyer function $\gm$ and is expressed as
\eq
\Phi_w =\arr{-\f{\sqrt{T}}{2}}{0}{0}{\f{\sqrt{T}}{2}}
\eqx 
The diagonal components of $\Phi_\wb$ become more complicated
\eq
\Phi_\wb=\arr{-\f{(e^{2\gm}+T\Tb e^{-2\gm})}{4\sqrt{T}}}{\f{(e^{2\gm}-T\Tb e^{-2\gm}) }{4\sqrt{T}}}{-\f{(e^{2\gm}-T\Tb e^{-2\gm}) }{4\sqrt{T}}}{\f{(e^{2\gm}+T\Tb e^{-2\gm}) }{4\sqrt{T}}}
=\scriptstyle
\arr{\scriptstyle-\f{1}{2} \sqrt{\Tb} \cosh \gmt}{\scriptstyle\f{1}{2} \sqrt{\Tb} \sinh \gmt}{\scriptstyle-\f{1}{2} \sqrt{\Tb} \sinh \gmt}{\scriptstyle\f{1}{2} \sqrt{\Tb} \cosh \gmt}
\eqx
however, the important observation made in \cite{AMSV} is that 
the diagonal components of
each expression can be treated as a single function defined on a double cover 
$\Sgt$ 
\eq
y^2=T(w)
\eqx
of the worldsheet $\Sg$.

In this manner one can rewrite the integral 
\eq
\int_{\Sigma} \left( 2\, \tr {\Phi_w \Phi_\wb}-\sqrt{T \Tb} \,d^2w\right)
\eqx
as an integral over $\Sgt$ of a wedge product of two \emph{closed} 1-forms:
\eq
\int_{\Sigma} \left( 2\, \tr {\Phi_w \Phi_\wb}-\sqrt{T \Tb} \,d^2w\right)=
\f{i}{2} \cdot \int_{\Sgt} \om \wedge \eta
\eqx
with
\eq
\om=\sqrt{T(w)} dw \qqqq 
\eta=\f{1}{2} \sqrt{\Tb(\wb)} \left( \cosh \gmt-1 \right) d\wb
+\f{1}{4} \f{1}{\sqrt{T(w)}} (\dw \gmt)^2 dw
\eqx
where for simplicity we used the original Pohlmeyer function (see (\ref{e.gmt})).
The $dw$ component of $\eta$ does not influence the integral but is chosen so that
$\eta$ is also closed ($d\eta=0$).

\begin{figure}
\hfill\includegraphics[height=7cm]{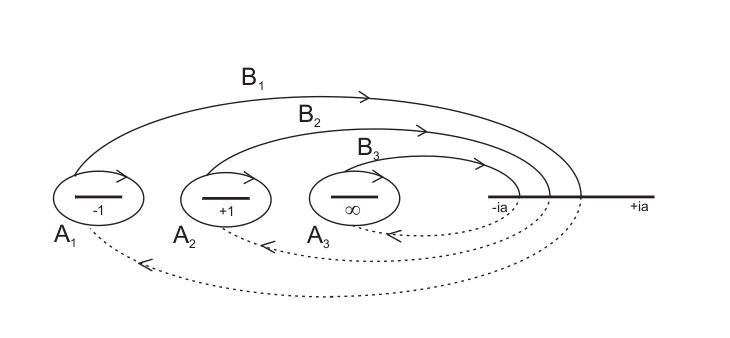}\hfill{}
\caption{Cycles on a genus 3 surface. For concrete computations it is 
convenient to make all the $B_i$ cycles to pass through the same point on
the last cut e.g. $w=0$.}
\label{f.genus3}
\end{figure}

If $\Sgt$ had genus $g$ (which is the generic case for Polygonal Wilson loops), 
one would use Riemann bilinear identity (or reciprocity)
to reduce the integral to products of integrals over cycles
\eq
\int_{\Sigma_g} \om \wedge \eta=\sum_{i=1}^g \int_{A_i} \om \int_{B_i} \eta - 
\int_{A_i} \eta \int_{B_i} \om \label{area genus}
\eqx
However in our case $\Sgt$ has genus 0, and the 1-forms may have singularities at
8 points (two copies of the 3 punctures and 2 branch points of the covering $y^2=T(w)$).
One possibility to proceed is to prove an analog of Riemann reciprocity 
directly for this case. The resulting expressions are, however, quite messy. 
In the end, we decided to adopt a slightly different strategy by 
treating the punctures
as infinitesimal branch cuts and using Riemann reciprocity for 
a genus 3 Riemann surface (see figure~\ref{f.genus3})
with an additional treatment of the singularities at the zeroes of $T(w)$.
Let us note that the $\eta$ 1-form is neither holomorphic or antiholomorphic and 
generic textbook formulas are not directly applicable.

\begin{figure}
\hfill\includegraphics[height=7cm]{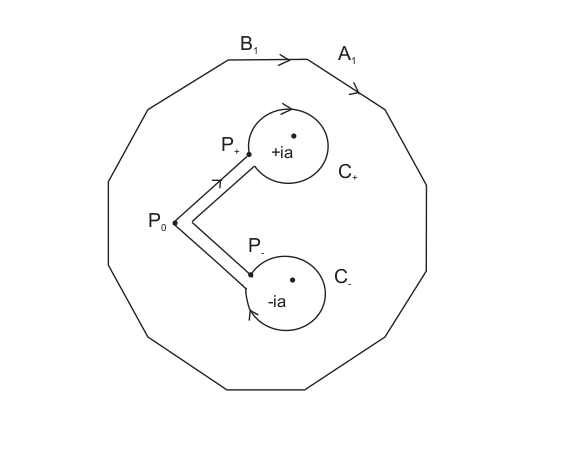} \hfill{}
\caption{The polygon is the standard representation of a genus 3 surface whose
boundary is the curve $L_{g=3}$ composed of the cycles $A_i$ and $B_i$ each 
traversed twice. The infinitesimal circles $C_\pm$ surround the singularities of 
$\eta$. $P_0$ is the (arbitrary but fixed) base point for constructing 
the function $F$ such that $\om=dF$.}
\label{f.recip}
\end{figure}

The idea of the derivation of the Riemann reciprocity formula is to rewrite
one of the forms as an exact form:
\eq
\om=dF
\eqx
where $F(P)=\int_{P_0}^P \omega$. This can always be done on the Riemann surface
minus some contour. Then one transforms the surface integral into a 1-dimensional
integral over that contour using Stokes theorem:
\eq
\int_{\Sgt} \om \wedge \eta =
\int_{\Sgt \setminus L} \om \wedge \eta = \int_{\Sgt \setminus L} d(F\eta)
=\int_L F\eta
\eqx
In our case, since $\om$ is regular\footnote{Recall that we are on $\Sgt$.} 
at the zeroes of $T(w)$, the contour $L$
may be taken to be the sum of the standard contour for a genus 3 surface 
$L_{g=3}$ and 
two infinitesimal circles $C_\pm$ around the zeros of $T(w)$ at $w=\pm i a$ 
as shown on figure~\ref{f.recip}.

Therefore
\eq
\label{area}
\int_{\Sgt} \om \wedge \eta = \int_{L_{g=3}} F\eta + \int_{C_{+}} F \eta + 
\int_{C_{-}} F \eta
\eqx
The first integral gives directly the standard bilinear expression
\eq
\label{periodsg3}
\int_{L_{g=3}} F\eta = \sum_{i=1}^3 \int_{A_i} \om \int_{B_i} \eta - 
\int_{A_i} \eta \int_{B_i} \om
\eqx
where we used the fact that
\eq
\int_{C_{+}} \om = \int_{C_{-}} \om =0
\eqx
Let us now concentrate on the remaining two terms. Now, 
\eqn 
\!\!\!\!\!\!\!\!\!\!\!\!\int_{C_{+}} F \eta &=& \int_{C_+} \left( \int_{P_0}^P \omega \right) \eta = \int_{C_+} \left[ \left( \int_{P_0}^{P_+} + \int_{P_+}^P \right) \omega \right] \eta \nonumber\\ 
&=&  \int_{P_0}^{P_+} \omega \int_{C_+} \eta + \int_{C_+} \left( \int_{P_+}^{P} \omega  \right) \eta =  \int_{P_0}^{P_+} \omega \int_{C_+} \eta - i\pi \frac{1}{6}
\eqnx
where the last integral is computed in Appendix~B.1.
Then, adding a similar expression for the second zero we get
\eqn
\left( \int_{C_{+}} + \int_{C_{-}} \right)F \eta  &=& \int_{P_0}^{P_+} \omega \int_{C_+} \eta +  \int_{P_0}^{P_-} \omega \int_{C_-} \eta  -2\cdot  i\pi \frac{1}{6} \nonumber\\
&=& \int_{P_-}^{P_+} \omega \int_{C_+} \eta -2\cdot  i\pi \frac{1}{6} \nonumber\\
&=& -2\cdot  i\pi \frac{1}{6}
\eqnx
where we  used the fact that the 1-form $\eta$ is regular everywhere apart 
from the zeros of $T(w)$ and so 
\eq
\int_{C_+} \eta + \int_{C_-} \eta =0
\eqx
Moreover, one can even show that $\int_{C_+} \eta =0$ (see Appendix~B.2). In this way we arrived at the final equality. 

Further, inserting this result into the integral (\ref{area}),
and computing the periods in (\ref{periodsg3}) using again
the regularity of $\eta$ outside the zeros of $T(w)$ and 
explicitly computing 
the integrals of the 1-form $\om$ 
\eq
\int_{A_1} \omega = \int_{A_2} \om = -2\pi i \frac{\Delta}{2}, \;\;\; \int_{A_3} \om = 2\pi i \frac{\Delta_{\infty}}{2}
\eqx 
we find that 
\eqn
\int_{\Sgt} \omega \wedge \eta  &=&  \sum_{i=1}^3 \int_{A_i} \omega \int_{B_i} \eta -2\cdot  i\pi \frac{1}{6}  \\
&=& 2\pi i \left[ -\frac{\Delta}{2}  \left( \int_{B_1} \eta+ \int_{B_2} \eta \right) +\frac{\Delta_{\infty}}{2} \int_{B_3} \eta -\frac{1}{6} \right] 
\eqnx 
The integrals over the cycles $B_i$ may be expressed by integrals between 
the punctures. From Fig.~\ref{f.genus3} and the antisymmetry of $\eta$ under 
changing of the Riemann sheet we find
\eq
2\int_{C_{-11}} \eta =  \int_{B_1} \eta + \int_{B_2} \eta, \;\;\; \int_{B_1} \eta = \int_{B_2}\eta,
\eqx
\eq 
2\int_{C_{1\infty}} \eta = \int_{B_1} \eta - \int_{B_3} \eta 
\eqx
Hence,
\eq
\int_{\Sgt} \omega \wedge \eta = -2 i \left[ \frac{\pi}{6} -\frac{\pi}{2} \left( (\Delta_{\infty} -2\Delta) \int_{C_{-11}} \eta- 2\Delta_{\infty} \int_{C_{1\infty}} \eta \right)\right] 
\eqx
Therefore the regularized Pohlmeyer contribution becomes
\eq
\label{e.regperiods}
\int_{\Sigma} \left( 2\, \tr {\Phi_w \Phi_\wb}-
\sqrt{T(w) \Tb(\wb)} \,d^2w\right) = \f{\pi}{6}-\f{\pi}{2} \left( (\Dl_\infty-2\Dl) 
\int_{C_{-1\,1}} \!\!\!\!\eta
-2\Dl_\infty \int_{C_{1\,\infty}} \!\!\!\!\eta \right)
\eqx
At this stage we have reduced the computation of the regularized Pohlmeyer contribution
to the evaluation of the integrals of the 1-form $\eta$ between the punctures.
This cannot be done directly, as we do not know of course the explicit form of
the Pohlmeyer solution $\gm$ or $\gmt$. However, as shown in \cite{AMSV}, 
the integrals of
$\eta$ can be related to the $\th \to -\infty$ ($\xi \to 0$) asymptotics of the
parallel transport of a solution along the curve which is a WKB line \cite{gaiotto}.

The main idea is to apply the well-know semiclassical methods where with the role of  the Plank constant is played by the spectral parameter $\xi$. Then, the linear problem 
\eq
(d+J)\Psi=0
\eqx
can be approximately solved with the leading contribution coming from the $\Phi_w$ part
\eq
\Psi \sim e^{\mp \frac{1}{2\xi} \int \sqrt{T(w)} dw}
\eqx
Clearly, the approximation is the best once we are on the WKB line defined as
\eq
\mbox{Im} \left( \frac{1}{\xi} \sqrt{T(w)} \dot{w} \right)=0
\eqx 
For our purposes, however, it is crucial to know also the subleading term related to the $\Phi_{\bar{w}}$ part of the flat connection 
\eq
e^{\pm \xi \int \tilde{\eta} } = e^{\pm \xi \int \left( \tilde{\eta} - \frac{1}{2} \sqrt{\bar{T}(\bar{w})} d\bar{w} \right)} e^{\pm \frac{\xi}{2} \int \sqrt{\bar{T}(\bar{w})} d\bar{w} }=e^{\pm \xi \int \eta  } e^{\pm \frac{\xi}{2} \int \sqrt{\bar{T}(\bar{w})} d\bar{w} }
\eqx
where $\tilde{\eta}$ is the 1-form $\eta$ without the subtraction term $\sqrt{\bar{T}(\bar{w})}/2$, exactly as it shows up in $\Phi_{\bar{w}}$.
\\
The basic object we want to compute in this limit is the skew product between two solutions at the punctures $j,k$. Then, the prescription is the following: \\
(i) take the known solution $\Psi_j(w_k')$ at $w=w_j'$ in the vicinity of the puncture $w_j$
\\
(ii) transport this solution via the parallel transport equation along a curve given by the WBK line equation to a point $w=w_k'$ near the puncture at $w_k$ taking into account the leading as well as the subleading terms
\\
(iii) compare the resulting with the known solution $\Psi_k(w_k')$ at $w=w_k'$.
\\
The resulting formula reads
\eqn
\lim_{\xi \rightarrow 0} \ss{j}{k} &=& e^{\frac{1}{\xi} \left[ \frac{1}{2}\int_{w_j'}^{w_k'} \sqrt{T(w)} dw + \frac{\Delta_j}{4} \log (w_j-w_j') + \frac{\Delta_k}{4} \log (w_k-w_k') \right] } \cdot \\
& &  e^{\xi \left[ \frac{1}{2}\int_{w_j'}^{w_k'} \sqrt{\bar{T}(\bar{w})} d\bar{w} + \frac{\Delta_j}{4} \log (\bar{w}_j-\bar{w}_j') + \frac{\Delta_k}{4} \log (\bar{w}_k-\bar{w}_k') \right] } \cdot e^{\xi \int_{w_j}^{w_k} \eta }
\eqnx
where the logarithmic terms are due to the exactly known form of the solution near the punctures. Moreover, these subtractions render the expression finite and therefore allow to extend the integration exactly to the punctures. This formula may be now compared with the exact expression for the skew product at any $\xi$ (\ref{e.s32}), which contains two undetermined zero mode constants $M,M^*$. Fortunately, they are given by the first two terms of the WKB approximation. Then, the path integral of $\eta$ may be given by a combination of the $\theta \rightarrow -\infty$ asymptotic of $f_a(\theta)$ function.

In this way we obtain the following explicit expressions for the period
integrals
\eqn
%h2[2dl-dlinf]+h2[2dl+dlinf]-2*h2[2*dl]
\int_{C_{-1\,1}} \eta &=& h(2\Dl-\Dl_\infty) + h(2\Dl+\Dl_\infty) -2h(2\Dl)
\nonumber \\
%h2[dlinf] + h2[2 dl + dlinf] - h2[2*dl] - h2[2*dlinf]
\int_{C_{1\,\infty}} \eta &=& h(\Dl_\infty)+h(2\Dl+\Dl_\infty) -h(2\Dl) -h(2\Dl_\infty)
\label{e.regperiods2}
\eqnx
where 
\eq
h(a)=\int_{-\infty}^{\infty} \frac{d\theta}{\pi} \cosh \theta \log \left(1-e^{-a\pi \cosh \theta} \right)
\eqx
Together with (\ref{e.regperiods}) it gives our final explicit expression for the
regularized Pohlmeyer contribution.

\subsection{Comparision with numerics}

Since the above derivation of (\ref{e.regperiods})-(\ref{e.regperiods2}) was 
quite complicated and involved many new ingredients, we decided to test the result
by numerically solving the modified sinh-Gordon equation on the 3-punctured
sphere and directly computing the regularized Pohlmeyer integral from the
numerical solution. We give some details on the numerical setup in Appendix A,
while here we just summarize the results and the comparision with the
analytical predictions (\ref{e.regperiods})-(\ref{e.regperiods2}).

\begin{figure}
\begin{minipage}[c]{0.45\textwidth}
\includegraphics[height=5cm]{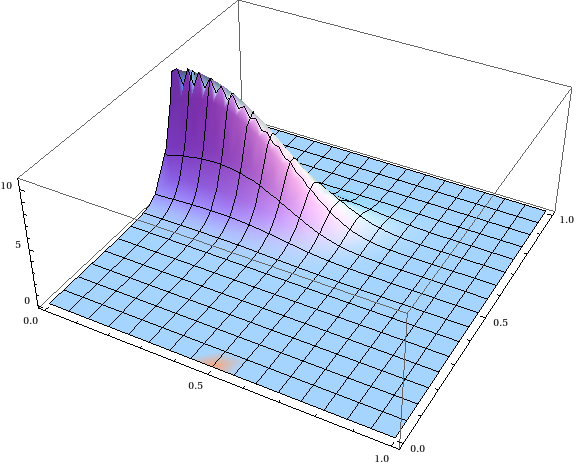} \hfill{}
\end{minipage}
\hfill
\begin{minipage}[c]{0.45\textwidth}

\begin{tabular}{|c|c|c|c|}
\hline
$\Dl$ & $\Dl_\infty$ & numerics & our formula\\
\hline
0.2& 0.3& 0.04536 & 0.0450779 \\
0.5& 0.9& 0.107649& 0.107622 \\
1.& 1.& 0.426311& 0.426166\\
1.& 1.05& 0.429572& 0.429503\\
2.& 2.& 0.517689& 0.517688\\
2.& 3.& 0.488985& 0.488985\\
4.& 4.& 0.523584& 0.523584\\
4.& 7.99& 0.0152435& 0.0152435 \\
\hline
\end{tabular}
\end{minipage}\hfill{}

\caption{Regularized action density for $\Dl=\Dl_\infty=4.0$ and a comparision
between the numerically evaluated regularized action and the analytical 
results of the formulas (\ref{e.regperiods})-(\ref{e.regperiods2}).}
\label{f.numerics}
\end{figure}

In figure~\ref{f.numerics} we show a plot of the integrand entering the
regularized Pohlmeyer action
\eq
\label{e.reg}
\int_{\Sigma} \left\{ \f{1}{2} \left( e^{2\gm(w,\wb)}+ T(w)\Tb(\wb) e^{-2\gm(w,\wb)} \right)
-\sqrt{T(w) \Tb(\wb)} \right\} d^2w
\eqx
in a qudrant of the (compactified) complex plane. The upper and right 
borders are mapped to
the puncture at $w=\infty$, while the puncture at $w=+1$ is right in 
the middle of the
lower border. The solution in the remaining three quadrants follows by symmetry.
In the table we have shown a comparision of the numerical evaluation of (\ref{e.reg})
together with the analytical results following from 
(\ref{e.regperiods})-(\ref{e.regperiods2}). The numerics becomes more difficult 
and less reliable for small $\Dl$'s. In particular the deviations in the first rows
of the table are within numerical errors estimated by changing the number of points of
the numerical grid. The remaining results show excellent agreement with the 
analytical formulas following from periods and solutions of the functional equations
for the overlaps.

\section{The regularized divergent contribution}

In this section we will deal with the remaining contribution to the action -- an
integral of $\sqrt{T\Tb}$ on the punctured sphere with specific cut-offs around
each puncture $|w-w_i|>\eps_i$. Since we adopt a similar regularization for the $S^5$
contribution (see (\ref{e.sv})), we have in fact two such contributions
\eq
\exp\left\{-\f{2\sqrt{\lm}}{\pi} \int_{\Sigma \setminus \{\eps_i\}} 
\sqrt{T(w)\Tb(\wb)}\, d^2w\right\}
\eqx
The above integral can be evaluated explicitly in the small $\eps_i$ limit 
relevant for us. To do it, we note that it can be expressed as an integral
of the wedge product of two closed 1-forms:
\eq
\int_{\Sigma \setminus \{\eps_i\}} \sqrt{T(w)\Tb(\wb)}\, d^2w=
const \cdot \int \sqrt{T} dw \wedge \sqrt{\Tb} d\wb
\eqx
and transformed into a product of residues of $\sqrt{T}$ times appropriate
(regularized) integrals over intervals between punctures.
The outcome is
\eq
\int_{\Sigma \setminus \{\eps_i\}} \sqrt{T(w)\Tb(\wb)}\, d^2w=Finite
-\f{\pi}{2} \Dl_\infty^2 \log\eps_\infty  -\f{\pi}{2} \Dl^2 \log \eps 
-\f{\pi}{2} \Dl^2 \log \eps 
\eqx
where
\eqn
\label{e.finite}
Finite&=&\f{\pi}{4} \Dl^2_\infty \biggl(2\log 2 +(1+a^2) \log 2+(-2-a^2+2\sqrt{1+a^2})
\log a^2 +\nonumber\\
&&+(1+a^2) \log(1+a^2)-4 \sqrt{1+a^2} \log(1+\sqrt{1+a^2}) \biggr)
\eqnx
Let us first concentrate on the logarithmically divergent part. Taking into account
the coupling-constant dependent prefactor, and generalizing slightly to three generic anomalous dimensions, we get
\eq
\exp\left\{ \sqrt{\lm} (\Dl_1^2 \log\eps_1+\Dl_2^2 \log \eps_2+
\Dl_3^2 \log \eps_3) \right\}
\eqx
We may use the relations (\ref{e.epsk}) to express the worldsheet cut-off in
terms of the physical target-space cut-off $z=\zeps$ and the products 
between the elementary solutions $1,2,3$ and one ($\Psi_A$) of the two 
solutions appearing in the reconstruction formulas 
of section~5. We get
\eq
\label{e.eqlogsprod}
\exp\left\{ \sum \DL_i \log \zeps+ \DL_1 \log |\ss{1}{A}|^2+ \DL_2 \log |\ss{2}{A}|^2+ 
\DL_3 \log |\ss{3}{A}|^2 \right\}
\eqx
where  $\DL_i\equiv \sqrt{\lm}\Dl_i$ is the unrescaled anomalous dimension.

We will now express the scalar products $\ss{k}{A}$ in terms of the gauge theory
operator insertion points $x_k$ and products between the elementary solutions.
The two solutions of the linear system $\Psi_A$ and $\Psi_B$ which determine
the target-space string embedding are completely specified by their coordinates
in e.g. the 1, $\bar{1}$ basis:
\eq
\Psi_A= \al 1+ \bt \bar{1} \qqqq \Psi_B= \gm 1+ \dl \bar{1}
\eqx 
where $\al \dl -\bt \gm=1$. Similarly we can express the 2 and 3 elementary 
solutions entering (\ref{e.eqlogsprod}) in terms of 1 and $\bar{1}$:
\eq
2=k 1+l \bar{1} \qqqq 3= m 1+n \bar{1}
\eqx
where $k$, $l$, $m$ and $n$ are appropriate overlaps evaluated at 
$\th=0$ ($\xi=1$). In terms of the above quantities, the part of (\ref{e.eqlogsprod})
depending on the products becomes
\eq
\sum_{k} \DL_k \log | \ss{k}{A} |^2 = \DL_1 \log \bt^2+ \DL_2 \log\, (k \bt-l\al)^2+
\DL_3 \log\,(m \bt-n \al)^2
\eqx
Now we may use formula (\ref{e.xk}) to relate the hitherto unknown coefficients
$\al$, $\bt$ and $\gm$ to the operator insertion points:
\eqn
x_1 &=& \f{\ss{1}{\Psi_B}}{\ss{1}{\Psi_A}} = \f{\dl}{\bt} \\
x_2 &=& \f{\ss{2}{\Psi_B}}{\ss{2}{\Psi_A}} = \f{k\dl-l\gm}{k\bt-l\al} \\
x_3 &=& \f{\ss{3}{\Psi_B}}{\ss{3}{\Psi_A}} = \f{m\dl-n\gm}{m\bt-n\al} 
\eqnx
Solving these equations with the constraint $\al \dl -\bt \gm=1$ yields
\eqn
\bt^2 &=& \f{ln}{lm-kn} \cdot \f{x_{23}}{x_{12} x_{13}} \\
(k \bt-l\al)^2 &=& \f{l}{n}(lm-kn) \cdot \f{x_{13}}{x_{12} x_{23}} \\
(m \bt-n \al)^2 &=& \f{n}{l} (lm-kn) \cdot \f{x_{12}}{x_{13} x_{23}} 
\eqnx
In the above formula $l=\ss{1}{2}_0 \equiv \ss{1}{2}_{\th=0}$, $n=\ss{1}{3}_0$,
while
\eq
lm-kn=\ss{1}{2}_0\sS{3}{1}_0-\sS{2}{1}_0\ss{1}{3}_0=\ss{3}{2}_0
\eqx 
using Schouten's identity.

Plugging the above expressions into (\ref{e.eqlogsprod}), we obtain finally
the standard CFT spacetime dependence of the 3-point function
\eq
\f{1}{\left( \f{x_{12}}{\zeps} \right)^{\DL_1+\DL_2-\DL_3}
\left( \f{x_{13}}{\zeps} \right)^{\DL_1+\DL_3-\DL_2}
\left( \f{x_{23}}{\zeps} \right)^{\DL_2+\DL_3-\DL_1}} \cdot ...
\eqx
multiplied by an additional contribution coming from the products of the
elementary solutions
\eq
\label{e.elem}
\exp^{\sqrt{\lm} \left( (\Dl_1+\Dl_2-\Dl_3) \log \ss{1}{2}_0 +
(\Dl_1+\Dl_3-\Dl_2) \log \ss{1}{3}_0 +
(\Dl_2+\Dl_3-\Dl_1) \log \ss{3}{2}_0 \right) }
\eqx
Going back to our solution of the functional equations (and returning to the
symmetric case of $\Dl_1=\Dl_2=\Dl$ and $\Dl_3=\Dl_\infty$) we see that the products
have the following structure at $\th=0$:
\eqn
\ss{1}{2}_0 &=& e^{M_{-11}+M_{-11}^*} \cdot e^{K_{-11}} \\
\ss{1}{3}_0=\ss{2}{3}_0 &=& e^{M_{1\infty}+M_{1\infty}^*} \cdot e^{K_{1\infty}} 
\eqnx
where 
\eqn
K_{-11} &=& k(2\Dl-\Dl_\infty) + k(2\Dl+\Dl_\infty) -2k(2\Dl) \\
K_{1\infty} &=& k(\Dl_\infty)+k(2\Dl+\Dl_\infty) -k(2\Dl) -k(2\Dl_\infty)
\eqnx
with
\eq
k(a)=\int_{-\infty}^\infty \f{d\th}{2\pi} \f{\log\left(1-e^{-a\pi \cosh \th}\right)}{\cosh \th}
\eqx
The zero-mode constants $M_{-11}$ and $M_{1\infty}$ are evaluated in Appendix C 
and after substituting into
(\ref{e.elem}), it turns out that they exactly \emph{cancel} the finite term
(\ref{e.finite}). Thus the remaining contribution to the OPE coefficient
becomes finally
\eq
\label{e.opediv}
\exp \left\{ \sqrt{\lm} \left[ (2\Dl-\Dl_\infty) K_{-11}+2\Dl_\infty K_{1\infty} \right]
\right\}
\eqx

\section{The final AdS contribution to the OPE coefficients}

We may now sum together the two contributions to the OPE coefficients -- 
(\ref{e.regperiods})-(\ref{e.regperiods2}) coming 
from the regularized Pohlmeyer integral and (\ref{e.opediv}) coming from the regularized
divergent integral. Both contributions have the same structure yielding
\eq
C^{OPE}_{AdS} = \exp\left\{-\f{\sqrt{\lm}}{6} -\sqrt{\lm} \left[ (2\Dl-\Dl_\infty) \tilde{P}_{-11}
+2\Dl_\infty \tilde{P}_{1\infty}\right]\right\}
\eqx 
with
\eqn
\tilde{P}_{-11} &=& \htl(2\Dl-\Dl_\infty) + \htl(2\Dl+\Dl_\infty) -2\htl(2\Dl) \\
\tilde{P}_{1\infty} &=& \htl(\Dl_\infty)+\htl(2\Dl+\Dl_\infty) -\htl(2\Dl) -\htl(2\Dl_\infty)
\eqnx
where
\eq
\htl(a)=\f{1}{2} h(a)-k(a)=
\f{1}{2\pi} \int_{-\infty}^\infty \f{\sinh^2\th}{\cosh\th} \log 
\left(1-e^{-a \pi \cosh \th} \right) d\th
\eqx

Several comments are in order here. Firstly, the above expression does not depend
on any details of the operators entering the OPE coefficient apart from their 
anomalous dimensions, thus it is universal for this class of operators. Secondly,
the above expression has to be supplanted by the regularized $S^5$ contribution 
\eq
C^{OPE}_{S^5} \equiv e^{-\f{\sqrt{\lm}}{\pi} \int_\Sigma \left( 
\text{$S^5$ contribution} -\sqrt{T\Tb} \right)}
\eqx
so one cannot draw conlusions on the behaviour of real OPE coefficients $C^{OPE}=
C^{OPE}_{AdS} \cdot C^{OPE}_{S^5}$, since the
latter part is currently unknown.
Thirdly, the factor $\exp(-\sqrt{\lm}/6)$ seems quite surprising, however its presence is essential
for sensible extremal and small $\Dl_i$ limits which we will examine shortly.

In this paper we have mostly considered the symmetric case of two equal anomalous dimensions. It should not be difficult to extend these considerations to the generic case of three distinct anomalous dimensions.
Repeating e.g. the analysis of the regularized Pohlmeyer contribution 
suggests the following structure.

Let us introduce the parameters $\al_i$:
\eq
\al_1=\Dl_2+\Dl_3-\Dl_1 \quad\quad\quad
\al_2=\Dl_1+\Dl_3-\Dl_2 \quad\quad\quad
\al_3=\Dl_1+\Dl_2-\Dl_3 
\eqx
Then the general answer should be
\eq
C^{OPE}_{AdS} =\exp \left\{ -\sqrt{\lm} \left( \f{1}{6} +F(\al_1,\al_2,\al_3)
\right) \right\}
\eqx
where
\eq
F(\alpha_1,\alpha_2,\alpha_3)=\alpha_1\htl (\alpha_1)+\alpha_2\htl (\alpha_2)+\alpha_3\htl (\alpha_3) +(\alpha_1+\alpha_2+\alpha_3) \htl (\alpha_1+\alpha_2+\alpha_3) \nonumber
\eqx
\eq
-(\alpha_1+\alpha_2) \htl (\alpha_1+\alpha_2)-(\alpha_1+\alpha_3) \htl (\alpha_1+\alpha_3)-(\alpha_3+\alpha_2) \htl (\alpha_3+\alpha_2)
\eqx
The structure of $F(\al_1,\al_2,\al_3)$ is very similar to the structure of formula
(7.11) in \cite{TristanKlose} but with a different function $\htl$ instead of
a logarithm. Below we will see that (7.11) arises from our formula in 
the limit of small anomalous dimensions.

\subsubsection*{Extremal limit}

In the extremal limit $\Dl_\infty=2\Dl$, all the terms with $\htl(a)$'s 
with nonzero arguments
will cancel between each other leaving the term $(2\Dl-\Dl_\infty) \htl(2\Dl-\Dl_\infty)$.
Now $\htl(a) \sim -\f{1}{6a}$ (see Appendix~D), so the remaining term will 
cancel with the 
$-\sqrt{\lm}/6$ giving the expected result
\eq
C^{OPE}_{AdS}(\Dl,\Dl,\Dl_\infty=2\Dl)=1
\eqx

\subsubsection*{Small $\Dl_i$ limit}

For small arguments, $\htl(a)$ behaves like (see Appendix~D)
\eq
\htl(a) \sim -\f{1}{6a} -\f{1}{2} \log a
\eqx
It turns out that contributions coming from the leading term will cancel out 
completely. The subleading logarithmic terms yield an expression
\eq
C^{OPE}_{AdS}(\Dl,\Dl,\Dl_\infty) \to
\left( \f{(2\Dl-\Dl_\infty)^{2\DL-\DL_\infty} (2\Dl+\Dl_\infty)^{2\DL+\DL_\infty}
\Dl_\infty^{2\DL_\infty} }{ (2\Dl)^{4\DL} (2\Dl_\infty)^{2\DL_\infty}} \right)^{\f{1}{2}}
\eqx
which coincides with formula (7.11) in \cite{TristanKlose}.

\subsubsection*{Large $\Dl_i$ limit and the Painleve transcendental}

For large arguments $\htl(a) \propto a^\# \cdot e^{-\pi a} $, and thus their contribution
is exponentially suppressed yielding a surprisingly simple universal limit independent
of the conformal dimensions of operators:
\eq
C^{OPE}_{AdS}(\Dl,\Dl,\Dl_\infty) \to \exp \left(-\f{\sqrt{\lm}}{6} \right)
\eqx
The simplicity of this result suggests that there should exist a much simpler
direct derivation of the above result\footnote{Thanks to Pedro Vieira for asking 
this interesting question.}. This turns out indeed to be the case.

In order to study the large $\Dl$ limit it is most convenient to study the
modified sinh-Gordon equation in its original formulation (\ref{e.pohl1eom})
\eq
\label{e.modshg}
\dw \dwb \gmt= \sqrt{T\Tb} \sinh \gmt
\eqx
where
\eq
T(w)= \f{\Dl_\infty^2}{4} \f{w^2+a^2}{(1-w^2)^2}
\eqx
The advantage of using (\ref{e.modshg}) is that $\gmt \to 0$ around the 
punctures. Hence it would seem naively that $\gmt=0$ would be a possible solution
of the equations of motion. This is not the case, however, as due to 
our genericity assumption on the nonvanishing of (\ref{e.pohl1}), 
$\gmt$ has to have logarithmic singularities
\eq
\label{e.logsing}
\gmt \sim \pm \log | w \pm i a|
\eqx
at the zeros of $T(w)$. Nevertheless, in the large $\Dl$ limit, 
when $\sqrt{T\Tb}$ is generically very large, in order to minimize
the string action (\ref{e.pohl1}), we expect to have an almost vanishing solution
with two narrow logarithmic spikes around the two zeros of $T(w)$.

Let us concentrate on the neighbourhood of $w=ia$ and introduce a new coordinate
through $w=u+i a$. Then (\ref{e.modshg}) takes the form
\eq
\dw \dwb \gmt = \underbrace{\f{\Dl_\infty^2}{4} \f{2a}{(1+a^2)^2}}_{C^2} \cdot 
\sqrt{u} \sqrt{\ub} \sinh \gmt
\eqx 
Redefining coordinates again
\eq
v=\f{2}{3} C u^{\f{3}{2}}
\eqx
yields the standard sinh-Gordon equation
\eq
\dw \dwb \gmt=\sinh \gmt
\eqx
In the large $\Dl$ limit the problem becomes rotationally invariant and 
after introducing a new variable $\gmt=2U$ and $R=2r \equiv 2|v|$ we obtain
the equation for a Painlev{\'e} III transcendent normalized as in \cite{ZamPainleve}:
\eq
U''+\f{1}{R} U'=\f{1}{2} \sinh 2U
\eqx
The coefficient of the logarithmic singularity (\ref{e.logsing}) becomes
\eq
U \sim  \pm \f{1}{3} \log R
\eqx
Now we may use the results of \cite{ZamPainleve} to evaluate directly 
the regularized Pohlmeyer action. We may rewrite the contribution around $w=ia$
as
\eq
\label{e.intcoshreg}
\int \sqrt{T\Tb} (\cosh \gmt -1) d^2w = \int_0^\infty (\cosh 2U-1) \f{3\pi R dR}{4}
\eqx
Note the $3\pi$ which comes from an angular integral corresponding to a $2\pi$ angle
in the original $u$ and $w$ coordinates. We then use the substitution 
from~\cite{ZamPainleve}
\eq
\f{1}{2} \cosh 2U =-\f{1}{R} \f{d}{dR} R F_c(R)
\eqx
to get
\eq
-\f{3\pi}{2}\int_0^\infty \left( \f{d}{dR} R F_c(R) +\f{R}{2} \right) dR
\eqx
This integral can be evaluated exactly using the asymptotic properties of 
$F_c(R)$ established in \cite{ZamPainleve}. At large $R$, $F_c(R) \sim -R/4$ 
up to exponentially small terms, while for small $R$, $R F_c(R) \to 1/18$. 
Thus the above
integral evaluates to $\pi/12$. Now taking into account two such contributions 
and the prefactor of the integral (\ref{e.intcoshreg}), we arrive directly at 
our universal large $\Dl$ limit:
\eq
e^{-\f{\sqrt{\lm}}{\pi} \left( \f{\pi}{12}+\f{\pi}{12} \right)} = 
e^{-\f{\sqrt{\lm}}{6}}
\eqx

\section{Summary and Outlook}

In this paper we have computed the universal part of the OPE coefficients
of three heavy operators with no Lorentz spins. This contribution comes from the
$AdS_2$ part of the string $\sg$-model, and has to be supplanted with the 
contribution of the $S^5$ part in order to obtain the full OPE coefficient
of the relevant operators.

We employed the methods of Pohlmeyer reduction, which have been previously
applied with great success to the case of null polygonal Wilson loops. It is interesting to notice that different aspects of the strong coupling physics of $\mathcal{N}=4$ SYM (gluon scattering amplitudes, anomalous dimensions through 2-point correlation functions, 
OPE coefficients) may be expressed by the same (modified) sinh-Gordon equation with all differences encoded in the analytical structure of the modification functions.  
Despite the similarities, the differences are significant, especially
in the analytical structure and target-space reconstruction (as we are 
dealing with $AdS_2$ instead of $AdS_3$) which make the
generalization nontrivial. As a cross-check of our results we made a comparison
of our formulas with direct numerical solution of the modified sinh-Gordon equation.

Unfortunately, as the OPE coefficients of operators dual to semiclassical spinning strings have not been previously calculated, in general there are no independent results allowing for testing our final expression. Even in the case were we have some information, like specific 
BPS operators with large charges, our lack of knowledge of the $S^5$ contribution
precludes a direct check. 

However, there are two limits in which we may cross-check our formula. First of all, in the extremal case we find that the AdS contribution to the OPE coefficient does not have
any semiclassical piece, as expected in this case. Secondly, when the anomalous dimensions are small (`medium'-type operators) our result is exactly equivalent to the Klose-McLoughlin formula obtained by a classical extremalization procedure of three point geodesics 
related to the three operators. It is reasonable to expect that such 
a point-like string/geodesic approximation to three point correlators is 
acceptable for not too `heavy' operators, which do not generate an extended surface.         

An additional consistency check is provided by the derivation of the correct 
CFT space-time dependence of the three-point correlators which arises from the
regularized divergent part.

It would be interesting to perform a comparision with the case of 
heavy-heavy-light correlators.
However it seems that such effects as leading order backreaction to the classical
2-point solution and corrections to vertex operators would have to be included 
in the heavy-heavy-light calculations in order for the comparision to be made.

There are several directions in which our work could be further developed. Obviously, as we consider entirely the AdS contribution to the OPE coefficient, one has to perform 
an analogous analysis for the $S^5$ part in order to obtain the full OPE coefficient
of the `heavy' operators. We expect that the $S^5$ contribution will depend on 
the particularities of the states in questions and not only on their conformal 
dimensions. Moreover, all selection rules should show up from the $S^5$ contribution. 

The $S^5$ part of the problem seems to be significantly more sophisticated 
for at least two reasons. 
The first reason is just technical, and should not pose too much problems.
Namely, the simplest string solution rotates in $S^3 \subset S^5$ which amounts to the reduction 
of the $\sigma$-model action to three dimensional case. Then, one should consider the 
corresponding Pohlmeyer reduction, again with a prescribed nonzero energy-momentum tensor. Since 
we deal with a higher dimensional target space, it leads to a more complicated (but still 
integrable) equations. 
A more serious problem is connected with the classical wavefunctions of the external states 
which must be taken into account as well. 
In particular it is not clear whether their contributions to the 3-point function
would cancel out with their contributions to the 2-point functions when constructing
a normalization independent OPE coefficient. Currently we lack an appropriate 
formulation with definite regularization prescription.      

Further generalization would involve the computation of the OPE coefficients of 
operators carrying charges related to $AdS_5$ momenta,
but here again a consistent treatment of the wavefunctions would be needed.
 
Finally, it would be very interesting to develop an analogous framework for 
higher point correlation functions and identify the general form of the key
functional equations for the overlaps in this case.

\bigskip

\noindent{\bf Acknowledgments:} We would like to thank Pedro Vieira and 
Amit Sever for initial collaboration, and Volodya Kazakov, Kostya Zarembo,
Kolya Gromov, Tristan McLoughlin and Arkady Tseytlin for interesting 
discussions. RJ was supported by Polish science funds as a research 
project N N202 105136 (2009-2012).

\appendix

\section{Details of the numerical computation}

In order to solve numerically the modified sinh-Gordon equation with the boundary
conditions relevant for 3-point correlation functions, we have to redefine the
Pohlmeyer function since both the original version $\gmt$ and the modified one $\gm$
are not suitable for numerics. $\gmt$ has logarithmic singularities at $w=\pm ia$,
while $\gmt$ blows up at the punctures. A convenient choice which is finite
everywhere is
\eq
\gm_3= \f{1}{2} \gmt +\f{1}{4} \log \left|w^2+a^2\right|^2 
-\f{1}{4} \log \left( |w|^4+a^2 \right)
\eqx
Another difficulty is to find a parametrization of the 3-punctured sphere which
would allow us to use a rectangular grid of finite dimension. By symmetry, one can
restrict oneself just to the upper quadrant $Re\, w>0$ and $Im\, w>0$.
In this quadrant we used the following mapping of $w=x+i y$:
\eq
s=\f{x}{x+1} \qqqq t=\f{y}{y+1}
\eqx
and considered the unit square $s\in [0,1]$ , $t \in [0,1]$. The boundary conditions
on the vertical edge $s=0$ follow from reflection symmetry, while those on
the horizontal edge $t=0$ follow from the condition that $\gmt$ vanishes on
the real line. This in turn is a consequence of the symmetry properties of the
target space solution under a reflection $y \to -y$ which ensures that 
$\partial_y x=\partial_y z=0$ at $y=0$.

The upper and right edges of the square get mapped to the puncture at $w=\infty$,
while $w=1$ is in the middle of the lower edge. We use Chebyshev spectral 
interpolation in the standard way in the $t$ coordinate and 
separately in \emph{two} subintervals $s \in [0,\f{1}{2}]$ and 
$s \in [\f{1}{2},1]$ in order to allow for nonanalyticity at the puncture $w=1$
(here $(s,t)=(\f{1}{2},0)$).

We use a Python interface to the PETSc library ({\tt petsc4py}) to solve
the spectrally discretized PDE. In order to get convergence we had to use
an automatic differentiation package (ADOL-C with Python bindings {\tt pyadolc}) 
to compute the Jacobian and use the LU linear solver from PETSc instead of 
the default iterative one.

\section{Evaluation of the integrals}

\subsection{$  \int_{C_+} \left( \int_{P_+}^{P} \omega  \right) \eta =  - i\pi \frac{1}{6}
$}

In the vicinity of the positive zero $w=ia$ the 1-forms in the leading order read
\eq
\omega \sim D \sqrt{w-ia}\; dw, \;\;\;\; \eta \sim - \frac{1}{16D} \frac{dw}{(w-ia)^2}
\eqx  
where $$ D=\frac{2ia \Delta_{\infty}^2}{4(1+a^2)^2}$$
Now, it is convenient to introduce a new variable $t^2=w-ia$. Since $p\in C_+$ i.e.,  is located on an infinitesimal circle around $w=ia$, we parametrize it as $t=\epsilon e^{i\phi}$, where $\epsilon$ is an infinitesimal parameter and $\phi$ the angular variable. Then, 
\eq
\int_{P_+}^P \omega = 2 \sqrt{D} \epsilon^3 \int _0^{\phi} e^{3i\phi'} i d\phi' +o(\epsilon^3)=   \frac{2}{3} \sqrt{D} \epsilon^3 (e^{3i\phi}-1) +o(\epsilon^3)
\eqx
Now, we may insert this result into the contour integral
\eq
 \int_{C_+} \left( \int_{P_+}^{P} \omega  \right) \eta = \int_0^{2\pi} \frac{2}{3} \sqrt{D} \epsilon^3 (e^{3i\phi}-1) \frac{-1}{16\sqrt{D}} \frac{1}{\epsilon^5 e^{5i\phi}} 2\epsilon^2e^{2i\phi} i d\phi +o(\epsilon^0)
 \eqx
 \eq
 = \frac{-i}{12} \int_0^{2\pi} (1-e^{-3i\phi})d\phi o(\epsilon^0)= \frac{-i\pi}{6}+o(\epsilon^0)
\eqx 

\subsection{$ \int_{C_{\pm}} \eta =0$}

Here we prove that the integral of $\eta$ around a zero of $T(w)$ vanishes. Without loosing generality we assume $a=1$ which fixes the position of zero $w_{\pm} = \pm i$. The cut is chosen to join the zeros. Now, consider an integral around $w_+$. The integral contour can be deformed in the following way
\begin{equation}
\int_{C_+} \eta = \left( \int_{x,-1_1} + \int_{-1_1,\infty_1} + \int_{\infty_1,1_1} +\int_{1_1,x} \right) \eta + \left( \int_{x,-1_2} + \int_{-1_2,\infty_2} + \int_{\infty_2,1_2} +\int_{1_2,x} \right) \eta 
\end{equation}
where the subscript denotes the Riemann sheet of the doouble cover $\Sgt$ 
and $x$ a fixed point on the cut. Now consider a single pair of integrals 
of the form
\eq
\int_{-1_1,\infty_1} \eta + \int_{-1_2,\infty_2} \eta
\eqx
Since $\eta$ is proportional to $\sqrt{\Tb}$, it differs on the two sheets
just by a relative sign. So the sum of the two integrals is zero. Repeating
this argument we arrive at
\eq
\int_{C_+} \eta =0
\eqx

\section{The zero mode parts}

In general, the zero-mode constants are given by the following expression
\eq
\label{e.M12}
M_{12} = \lim_{w_1' \rightarrow w_1} \lim_{w_2' \rightarrow w_2}  \int_{w_1'}^{w_2'} \frac{1}{2} \sqrt{T(w)} dw + \frac{\Delta_1}{4} \ln (w_1-w_1') +\frac{\Delta_2}{4} \ln (w_2-w_2')
\eqx 
where $w_1,w_2$ are the end points for the WKB trajectory (poles of $T(w)$) with corresponding anomalous dimensions $\Delta_1, \Delta_2$.  In our case 
\begin{equation*}
\textstyle
\int \sqrt{T(w)} dw = \frac{\Delta_{\infty}}{2} \int \frac{\sqrt{w^2+a^2}}{1-w^2} dw =
\end{equation*}
\begin{equation*}
\textstyle
=
\frac{\Delta_{\infty}}{4} \left[ \sqrt{1+a^2} \ln \left( \frac{1+w}{1-w} \cdot \frac{a^2+w+\sqrt{1+a^2} \sqrt{a^2+w^2}}{a^2-w+\sqrt{1+a^2} \sqrt{a^2+w^2}} \right)   - 2\ln (w+\sqrt{a^2+w^2})\right]
\end{equation*}
\eq
\textstyle
= \frac{\Delta}{2} \cdot  \ln \left( \frac{1+w}{1-w} \cdot \frac{a^2+w+\sqrt{1+a^2} \sqrt{a^2+w^2}}{a^2-w+\sqrt{1+a^2} \sqrt{a^2+w^2}} \right)   -  \frac{\Delta_{\infty}}{2} \cdot \ln (w+\sqrt{a^2+w^2})
\eqx
One may easily verify that the first term gives the expected 
asymptotics at $w=1$ and $w=-1$ (with the correct $\frac{\Delta}{2}$ factor 
due to the fact that $\sqrt{1+a^2}=\frac{2\Delta}{\Delta_{\infty}}$), while the second term is responsible for the asymptotics at $w=\infty$.

For the punctures $w_1=-1, w_2=1$ we get a cancellation of singularities 
in (\ref{e.M12}) and find the following finite result 
\eq
M_{-1,1} = \frac{\Delta}{2} \ln \frac{8\Delta^2}{4\Delta^2-\Delta^2_{\infty}} - \frac{\Delta_{\infty}}{4} \ln \frac{2\Delta +\Delta_{\infty}}{2\Delta - \Delta_{\infty}} + \frac{\Delta}{4} i\pi
\eqx
In order to calculate $M_{\infty,1}$ it is convenient to perform a change of variables $w \rightarrow \frac{1}{z}$. Then $M_{\infty,1} = \tilde{M}_{0,1}$, where
\eq
\tilde{M}_{0,1} = \lim_{z_1' \rightarrow z_1=0} \lim_{z_2' \rightarrow z_2=1}  \int_{z_1'}^{z_2'} \frac{1}{2} \sqrt{T (1/z)} \frac{dz}{-z^2} + \frac{\Delta_{\infty}}{4} \ln (z_1-z_1') +\frac{\Delta}{4} \ln (z_2-z_2')
\eqx
Therefore, 
\eq
\textstyle
\left. \tilde{M}_{0,1} = \lim_{z_1' \rightarrow z_1=0} \lim_{z_2' \rightarrow z_2=1} \frac{\Delta}{4} \cdot  \ln \left( \frac{1+\frac{1}{z}}{1-\frac{1}{z}} \cdot \frac{a^2+\frac{1}{z}+\sqrt{1+a^2} \sqrt{a^2+\frac{1}{z^2}}}{a^2-\frac{1}{z}+\sqrt{1+a^2} \sqrt{a^2+\frac{1}{z^2}}} \right)  \right|_{z_1'}^{z_2'}
\eqx
\begin{equation*}
\textstyle
\left.  -  \frac{\Delta_{\infty}}{4} \cdot \ln \left(\frac{1}{z}+\sqrt{a^2+\frac{1}{z^2}} \right)  \right|_{z_1'}^{z_2'}  + \frac{\Delta_{\infty}}{4} \ln (-z_1') +\frac{\Delta}{4} \ln (z_2-z_2') 
\end{equation*}
The singular part at $z_1' \rightarrow z_1=0$ is
\eq
\textstyle
 \frac{\Delta_{\infty}}{4} \cdot \ln \left(\frac{1}{z_1'}+\sqrt{a^2+\frac{1}{z_1'^2}} \right) + \frac{\Delta_{\infty}}{4} \ln (-z_1') =   \frac{\Delta_{\infty}}{4} \cdot \ln \left(\frac{2}{z_1'} \right) + \frac{\Delta_{\infty}}{4} \ln (-z_1')  =  \frac{\Delta_{\infty}}{4} (\ln 2 + i \pi)
 \eqx
while the finite part at $z_1'\rightarrow z_1=0$ reads
\eq
\scriptstyle
 \lim_{z_1' \rightarrow z_1=0} - \frac{\Delta}{4} \cdot  \ln \left( \frac{1+\frac{1}{z_1'}}{1-\frac{1}{z_1'}} \cdot \frac{a^2+\frac{1}{z_1'}+\sqrt{1+a^2} \sqrt{a^2+\frac{1}{z_1'^2}}}{a^2-\frac{1}{z_1'}+\sqrt{1+a^2} \sqrt{a^2+\frac{1}{z_1'^2}}} \right) = -\frac{\Delta}{4} \ln \left( \frac{1+\sqrt{1+a^2}}{-1+\sqrt{1+a^2}}\right) -\frac{\Delta}{4} \ln(-1)
\eqx
The part at $z_1' \rightarrow z_1=1$ is
\begin{equation*}
\scriptstyle
 \lim_{z_2' \rightarrow z_2=1}  \frac{\Delta}{4} \cdot  \ln \left( \frac{1+\frac{1}{z_2'}}{1-\frac{1}{z_2'}} \cdot \frac{a^2+\frac{1}{z_2'}+\sqrt{1+a^2} \sqrt{a^2+\frac{1}{z_2'^2}}}{a^2-\frac{1}{z_2'}+\sqrt{1+a^2} \sqrt{a^2+\frac{1}{z_2'^2}}} \right)-\frac{\Delta_{\infty}}{4} \cdot \ln \left(\frac{1}{z_2'}+\sqrt{a^2+\frac{1}{z_2'^2}} \right) +\frac{\Delta}{4} \ln (1-z_2') =
\end{equation*}
\eq
\scriptstyle
\hspace*{-0.7cm} \lim_{z_2' \rightarrow z_2=1}  \frac{\Delta}{4} \cdot  \ln \left( \frac{z_2'+1}{z_2'-1} \cdot \frac{a^2+\frac{1}{z_2'}+\sqrt{1+a^2} \sqrt{a^2+\frac{1}{z_2'^2}}}{a^2-\frac{1}{z_2'}+\sqrt{1+a^2} \sqrt{a^2+\frac{1}{z_2'^2}}} \right)-\frac{\Delta_{\infty}}{4} \cdot \ln \left(\frac{1}{z_2'}+\sqrt{a^2+\frac{1}{z_2'^2}} \right) +\frac{\Delta}{4} \ln (z_2'-1) + \frac{\Delta}{4}i\pi 
\eqx
The singular parts $\sim \ln (z_2'-1)$ again cancel and we are left with
\eq
\textstyle
 \frac{\Delta}{4}  \ln \frac{2(a^2+1)}{a^2}-\frac{\Delta_{\infty}}{4} \cdot \ln \left(1+\sqrt{a^2+1} \right)+ \frac{\Delta}{4}i\pi 
\eqx
Finally we get
\eq
\textstyle
\tilde{M}_{0,1} = \frac{\Delta_{\infty}}{4} (\ln 2 + i \pi) -\frac{\Delta}{4} \ln \left( \frac{1+\sqrt{1+a^2}}{-1+\sqrt{1+a^2}}\right)
+ \frac{\Delta}{4}  \ln \frac{2(a^2+1)}{a^2}-\frac{\Delta_{\infty}}{4} \cdot \ln \left(1+\sqrt{a^2+1} \right)\eqx
which may be simplified to (recall that $M_{\infty,1} = \tilde{M}_{0,1}$)
\eq
\textstyle
M_{\infty,1} = \frac{\Delta_{\infty}}{4} (\ln 2 + i \pi)+ \frac{\Delta}{4}  \ln 2-\frac{\Delta_{\infty}}{4} \cdot \ln \left(\frac{\Delta_{\infty}+2\Delta}{\Delta_{\infty}} \right) -\frac{\Delta}{2} \ln \left( \frac{2\Delta +\Delta_{\infty}}{2\Delta}\right)
\eqx

\section{Small $a$ asymptotics of $\tilde{h}(a)$}

We divide the integral into two parts
\eq
\tilde{h} (a) =  \frac{1}{2\pi} \int_{-\infty}^{\infty} \frac{\sinh^2 \theta }{\cosh \theta} \ln (1-e^{-a\pi \cosh \theta}) 
\eqx
\eq
= \frac{1}{\pi} \left[  \int_{0}^{\infty} \cosh \theta \ln (1-e^{-a\pi \cosh \theta})  -  \int_{0}^{\infty} \frac{1 }{\cosh \theta} \ln (1-e^{-a\pi \cosh \theta})\right]
\eqx
The first integral was calculated in \cite{Didina} and has 
the following asymptotics for small $a$
\eq
 \int_{0}^{\infty} \cosh \theta \ln (1-e^{-a\pi \cosh \theta}) = -\frac{\pi}{6a} + o(1)
\eqx
In order to compute the second integral we notice that
\eq
\ln (1-e^{-a\pi \cosh \theta} ) = \ln (a \cosh \theta) + o(1)
\eqx  
Then, the integral is finite in this limit and reads 
\eqn
 \int_{0}^{\infty} \frac{1 }{\cosh \theta} \ln (1-e^{-a\pi \cosh \theta}) &=& \int_{0}^{\infty} \frac{1 }{\cosh \theta}\ln (a \cosh \theta) +o(1)\\
 &=& \frac{\pi}{2} \ln a +o(1)
 \eqnx
Inserting these results to the integral we get at the leading order
\eq
\tilde{h}=- \frac{\pi}{6a}-\frac{1}{2}\ln a 
\eqx

\end{document}